\documentclass[letter]{aa} 
\usepackage[varg]{txfonts}
\usepackage{natbib}
\usepackage{graphicx}
\usepackage{txfonts}
\usepackage{tablefootnote}
\def\hii{\mbox{H\,{\sc ii}}}
\begin{document}

   \title{A CO Funnel in the Galactic Centre: Molecular Counterpart of the Northern Galactic Chimney?}

   \author{V. S. Veena\inst{1,2} \and
          D. Riquelme\inst{1,3,4} \and
          W.-J. Kim\inst{2} \and
          K. M. Menten\inst{1} \and
          P. Schilke\inst{2} \and
          M. C. Sormani\inst{5} \and
          W. E. Banda-Barrag\'an\inst{6,7} \and
          F. Wyrowski\inst{1} \and
          G. A. Fuller\inst{2,8} \and
          A. Cheema\inst{1}
}
   \institute{Max-Planck-Institut f\"ur Radioastronomie, auf dem H\"ugel 69, 53121 Bonn, Germany\\
              \email{vadamattom@mpifr-bonn.mpg.de}
    \and 
    I. Physikalisches Institut, Universit\"at zu Köln, Z\"ulpicher Str. 77, 50937 K\"oln, Germany
    \and
    Instituto Multidisciplinario de Investigaci\'on y Postgrado, Universidad de La Serena, Ra\'ul Bitr\'an 1305, La Serena, Chile
    \and
    Departamento de Astronom\'ia, Universidad de La Serena, Av. Cisternas 1200, La Serena, Chile
    \and
    Institut f\"ur Theoretische Astrophysik, Zentrum f\"ur Astronomie, Universit\"at Heidelberg, Albert-Ueberle-Str. 2, 69120 Heidelberg, Germany
    \and
    Escuela de Ciencias F\'isicas y Nanotecnolog\'ia, Universidad Yachay Tech, Hacienda San Jos\'e S/N, 100119 Urcuqu\'i, Ecuador
    \and
    Hamburger Sternwarte, University of Hamburg, Gojenbergsweg 112, 21029 Hamburg, Germany
    \and
    Jodrell Bank Centre for Astrophysics, Department of Physics and Astronomy, The University of Manchester, Manchester M13 9PL, UK
 }
 
    \date{Received ; accepted }

 
\abstract
   {We report the discovery of a velocity coherent, funnel shaped $^{13}$CO emission feature in the Galactic centre (GC) using data from the SEDIGISM survey. The molecular cloud appears as a low velocity structure (V$_\textrm{LSR}$=[--3.5, $+$3.5] km~s$^{-1}$) with an angular extent of $0.95^\circ\times1^\circ$, extending toward positive Galactic latitudes. The structure is offset from Sgr A$^*$ toward negative Galactic longitudes and spatially and morphologically correlates well with the northern lobe of the 430 pc radio bubble, believed to be the radio counterpart of the multiwavelength GC chimney. Spectral line observations in the frequency range of 85--116 GHz have been carried out using the IRAM 30 metre telescope toward 12 positions along the funnel-shaped emission. We examine the $^{12}$C/$^{13}$C isotopic ratios using various molecules and their isotopologues. The mean $^{12}$C/$^{13}$C isotope ratio ($30.6\pm2.9$) is consistent with the structure located within inner 3 kpc of the Galaxy and possibly in the GC. The velocity of the molecular funnel is consistent with previous radio recombination line measurements of the northern lobe of radio bubble. Our multiwavelength analysis suggests that the funnel shaped structure extending over 100 pc above the Galactic plane is the molecular counterpart of the northern GC chimney. }


   \maketitle
%

\section{Introduction}
High resolution observations of external galaxies unveil outflows from galactic nuclei that are driven either by a nuclear starburst or an active galactic nucleus \citep[][]{{2014A&A...562A..21C},{2016A&A...588A..41C}}. These outflows extend from dense star forming regions near the galactic mid-planes up to several kilo parsecs away from them and are capable of ejecting ionised, atomic and molecular gas from their host galaxies \citep{2017ARA&A..55..389T}. Hence, they are often considered a negative feedback process that contributes to the star formation quenching of a galaxy \citep[][]{{2012ARA&A..50..455F},{2014Natur.516...68G}}. Conversely, these outflows could constitute a positive trigger, with compression of gas within the outflows leading to gravitational collapse and subsequent star formation \citep[][]{{2013MNRAS.433.3079Z},{2016MNRAS.455.4166B}}.
\begin{figure*}[!htbp]
\centering
\hspace*{-0.4cm}
\includegraphics[scale=0.82]{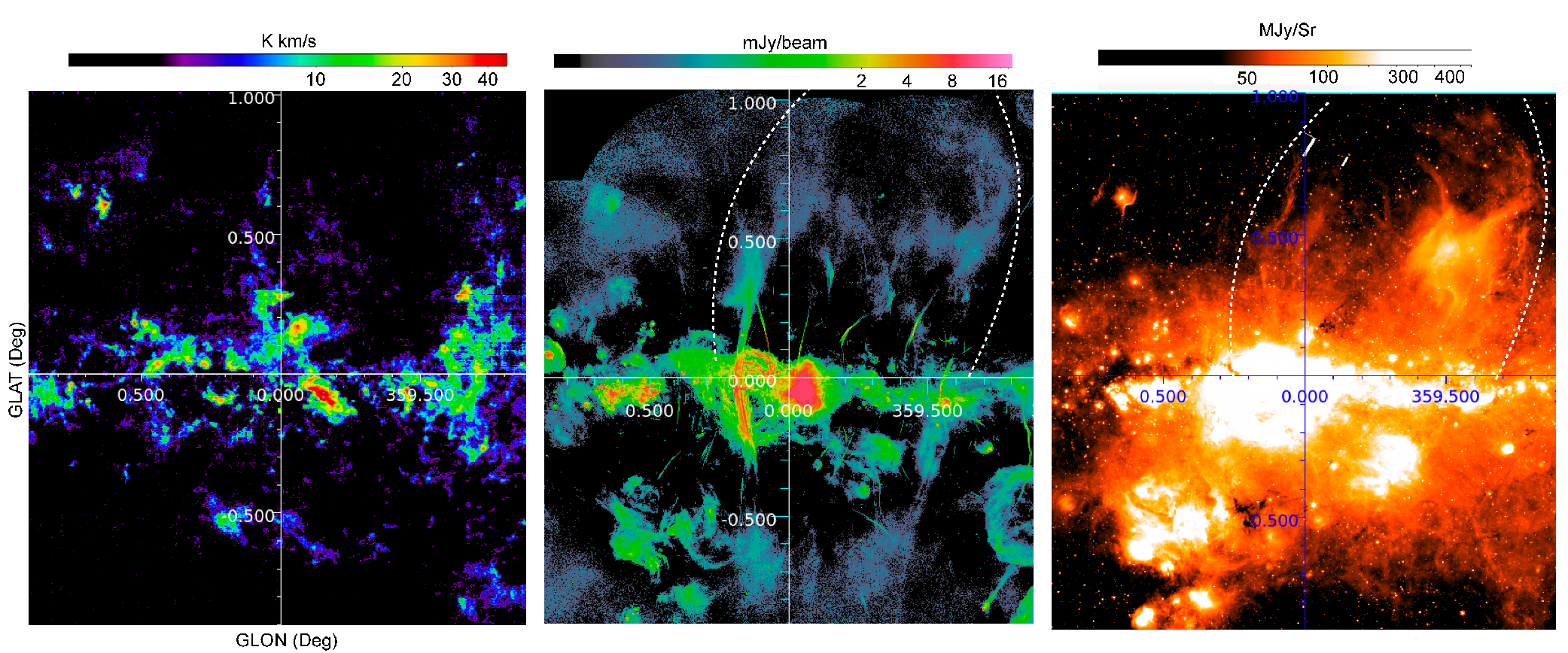}
\caption{Multiwavelength view of the GC chimney.  (\textit{Left}) SEDIGISM $^{13}$CO velocity integrated intensity map in the velocity range from --3.5 to 3.5 km~s$^{-1}$ showing the funnel shaped CO feature we name the Galactic Centre Molecular Funnel or GCMF, proposed to be the molecular counterpart of the radio bubble. (\textit{Middle}) MeerKAT 1~GHz radio emission map toward the GC region \citep{2022ApJ...925..165H} showing the 430 pc bipolar radio bubble (Heywood et al.,2019), (\textit{Right}) Spitzer 24~$\mu$m warm dust emission. The IR counterpart of the northern lobe of the radio bubble (roughly outlined as white dashed curves in middle and right panels) is seen as filamentary features extending to northern latitudes.}
\label{GCbub}
\end{figure*}
Radio observations of the Galactic centre (GC) region in the 1980s revealed evidence of a large scale outflow from the GC.  Emission extending several hundreds of parsecs was identified \citep[e.g.,][]{{1972ApJ...175L.127S},{1972NPhS..238..105K},{1989ApJ...341L..47S}} that was believed to originate from the high energy activity in the core of the Milky Way.  This was later confirmed with the discovery of the \textit{Fermi} bubbles,  which comprise of bipolar lobes extending vertically up to 8 kpc, observed in the $\gamma$-ray bands \citep{2010ApJ...724.1044S}, which also have  radio counterparts \citep{2013Natur.493...66C}. \citet{2020Natur.588..227P} reported the discovery of soft-X-ray emitting bubbles extending 14 kpc above and below the GC that are believed to originate in large energy injections from the GC. Other studies of the GC in X-ray bands uncovered two oppositely directed, 200-pc chimneys of hot plasma at the base of the \textit{Fermi} bubbles \citep{{2003ApJ...582..246B},{2013ApJ...773...20N}}.  Recently, \citet{2019Natur.573..235H} reported the discovery of  radio counterparts at 1 GHz of these chimneys, extending to 430 pc in the vertical plane. \citet{2021A&A...646A..66P} examined the spatial relationship between the emission observed in different wavebands. Based on H$_2$ column density maps derived from  $Herschel$ dust continuum data, they also identified molecular gas within few tenths of parsecs above the GC that is interpreted as molecular gas outflowing within the chimneys. Their analysis revealed that the X-ray, radio and infrared emissions are deeply interconnected, forming coherent features on scales of hundreds of parsecs, representing the channel connecting the quasi-continuous, intermittent activity at the GC within the base of the \textit{Fermi} bubbles. However, there are also studies that question the association of this structure with the GC. For example, \citet{2020PASJ...72L..10T} suggested that the western part of the northern lobe is an \hii~region located within a few kpc from us (at a heliocentric distance of $\sim2$ kpc) and not in the GC region.\\
In this letter we report the discovery of a prominent funnel shaped CO emission feature in the GC extending spatially over 100 pc, both in latitude as well as longitude. We investigate the possible association of this cloud with the GC chimneys using CO data from APEX 12 m telescope, 3~mm molecular line data from IRAM 30 m telescope and multiwavelength archival data. The details of our observations are described in Section 2. The results are presented in Sections 3 and 4 and discussed in Section 5. Finally in Section 6, we summarise our conclusions.
\section{Observations and Archival Data}
\subsection{The SEDIGISM survey}
The Structure, Excitation and Dynamics of the Inner Galactic Interstellar Medium \citep[SEDIGISM;][]{{2017A&A...601A.124S},{2021MNRAS.500.3064S}} is a survey that covers a 84 deg$^2$ area of the inner Galactic plane ($-60^\circ<l<+31^\circ$, $|b| \le 0\rlap{.}^\circ$5) in the $J=2-1$ transition of $^{13}$CO and C$^{18}$O isotopologues of the carbon monoxide molecule, which probe the moderately dense (n(H$_2)\sim10^3$~cm$^{-3}$) component of the interstellar medium (ISM). The coverage is extended in latitude to $|b| \le 1\rlap{.}^\circ$0 around the GC ($-2^\circ<l<+1.5^\circ$). The observations were carried out with  Swedish Heterodyne Facility Instrument (SHFI) at the Atacama Pathfinder Experiment 12 metre submillimetre telescope \citep[APEX;][]{2006A&A...454L..13G}. The spectral and angular resolutions are 0.25 km~s$^{-1}$ and 30$\arcsec$, respectively, the  pixel size is 9.5$\arcsec$ and the rms sensitivity is 0.9 K in main-beam brightness temperature units, T$_\textrm{MB}$.
\subsection{IRAM 3 mm Observations}
Observations of spectral lines in the 3 mm wavelength range were carried out using the Institut de Radioastronomie Millim\'etrique (IRAM) 30 metre telescope during the period 2022 December 24--26 (Project ID: 144-22) with the Eight MIxer Receiver \citep[EMIR;][]{2012A&A...538A..89C}. The position switching On--Off mode observations were performed with two frequency setups covering the range 84.9--116.7 GHz that allowed us to observe transitions of various molecular species. We used the FTS200 backend with a spectral resolution of 200 kHz (equivalent to 0.6 km~s$^{-1}$ at 3 mm) and the data reduction was performed using the Continuum and Line Analysis Single-dish Software (CLASS) which is part of the GILDAS software \citep{pety2005_gildas}. 
\section{An Extended CO Funnel-like Structure Near SgrA$^*$}
\begin{figure}
\centering
\hspace*{-0.3cm}
\includegraphics[scale=0.67]{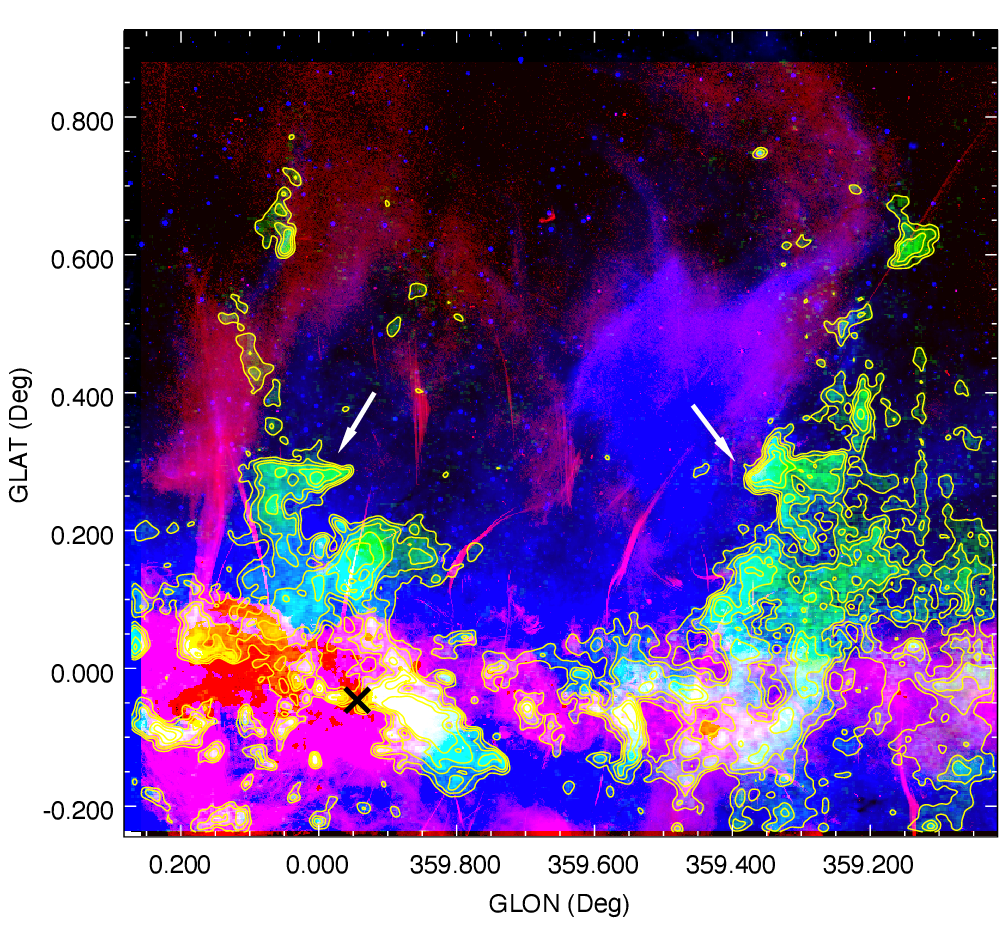}  
\caption{Three color composite image showing the MeerKAT 1~GHz radio emission (red), SEDIGISM $^{13}$CO emission (green) and Spitzer 24~$\mu$m emission (blue). The $^{13}$CO emission contours are shown in yellow. The contour levels are 4, 6.5, 9, 14 and 20 K\,kms$^{-1}$. The position of Sgr A$^*$ is marked as $\times$. White arrows point to finger-like protrusions.}
\label{rgb}
\end{figure}

In order to search for a possible molecular counterpart of the northern Galactic chimney, we examined the $^{13}$CO (2--1) data from the SEDIGISM survey. In channel maps of the SEDIGISM data, we identify a prominent shell-like feature centred at $l\sim359.6^\circ$, within the GC region in the velocity range from $-3.5$ km~s$^{-1}$ to +3.5 km~s$^{-1}$ (see Fig.~\ref{channel}). Fig.~\ref{GCbub}(\textit{Left}) shows the velocity integrated intensity map over the same velocity range. The emission resembles a funnel with an angular extent of $0.95^\circ\times1^\circ$, narrower close to the Galactic plane (0.2$^\circ$ at $b\sim-0.2^\circ$) and wider at higher latitudes (0.9$^\circ$ at $b\sim0.8^\circ$).  The centroid of this structure is offset  from Sgr A$^*$. We call this feature the Galactic Centre Molecular Funnel (GCMF).   
In Fig.~\ref{GCbub}(\textit{Middle}), we present the MeerKAT 1~GHz radio image from \citet{2019Natur.573..235H}, which shows a 430 pc bipolar radio bubble associated with the GC chimney. The northern lobe of the bipolar bubble is asymmetric with respect to Sgr A$^*$, shifted westwards, toward the fourth Galactic quadrant.  Fig.~\ref{GCbub}(\textit{Right}) shows the Spitzer 24~$\mu$m mid-infrared (MIR) image of the GC where the MIR counterpart of the northern radio bubble is seen as filamentary structures extending up to 1$^\circ$ in latitude. Comparison of the GCMF with radio and MIR emission reveals that $^{13}$CO emission is spatially and morphologically well correlated with the northern lobe of the GC chimney. This is evident from Fig.~\ref{rgb} as well as Fig.~\ref{funnelinfra}. Finger-like protrusions are found around $b\sim0.3^\circ$, pointing radially inward toward the interior of the funnel/shell. Such fingers are observed in star forming regions and Galactic supershells \citep[e.g.,][]{2013PASA...30...25D}. They are suggested to originate from the interaction of the expanding shock with pre-existing molecular clouds, caused by the  Rayleigh-Taylor instability on the inner surface of the supershells. The resultant clumplets exposed to the hot high-velocity winds get subsequently evaporated and acquire a head-tail structure \citep{2015ApJ...799...64D}.
\section{Molecular Line Survey of the GCMF}
\begin{figure}[!tbp]
\centering
\includegraphics[scale=0.48]{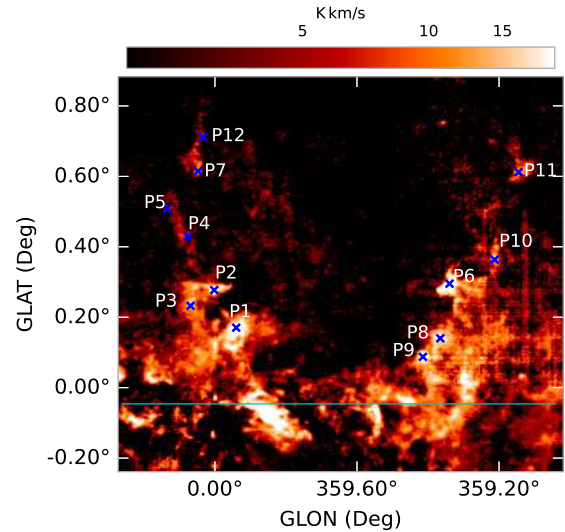}  
\caption{$^{13}$CO integrated intensity map. Twelve positions used for IRAM 3 mm line survey are marked. Cyan line shows the physical mid-plane of the Galaxy.}
\label{13coiso}
\end{figure}

To investigate the chemical properties of the GCMF, we carried out a molecular line survey of the GCMF at 3 mm using IRAM 30 m telescope. For this, we selected  12 positions along the GCMF (see Fig.~\ref{13coiso}), targeting clumps of high $^{13}$CO emission intensity. We only considered positions above $b>0.08^\circ$ to avoid strong contamination from the Galactic mid-plane. The observations reveal the rich chemistry of the GCMF. We present below some key first results, while a more comprehensive analysis will be presented in a future paper. 
\subsection{$^{12}$C/$^{13}$C isotope ratio}
Important information on the  chemical evolution of 
galaxies can be inferred from elemental isotopic abundance ratios. In particular, $^{12}$C/$^{13}$C ratio is an excellent probe for such studies as $^{12}$C is predicted to form predominantly in triple alpha reaction in massive stars \citep{1995ApJS...98..617T} whereas $^{13}$C is a product of secondary processes such as CNO-I cycle (in intermediate and massive stars), burning of $^{12}$C in low metallicity massive stars and proton-capture nucleosynthesis in AGB stars \citep{{1994ARA&A..32..153M},{1999A&A...346L..37L},{2002A&A...390..561M}}. Various studies have  investigated variation of $^{12}$C/$^{13}$C across the Galactic disk \citep{{1990ApJ...357..477L},{2005ApJ...634.1126M},{Jacob2020},{2023A&A...670A..98Y}}. One common method is to measure the intensity ratio of $^{12}$C and $^{13}$C isotopologues of a molecule. A disadvantage is that the emission from more abundant ($^{12}$C-bearing) species could be saturated, leading to underestimated isotope ratios. Another effect is chemical fractionation in which one isotope could be preferentially enriched compared to the other under certain conditions \citep{{1984ApJ...277..581L},{2020A&A...640A..51C},{Jacob2020}}. Nevertheless, $^{12}$C/$^{13}$C studies of Galactic clouds reveal that a systematic gradient exists in the Milky Way with the $^{12}$C/$^{13}$C ratio steadily increasing with distance from the GC \citep[e.g.,][]{2014A&A...570A..65G}. 
\begin{figure*}[!htbp]
\centering
\includegraphics[scale=0.45]{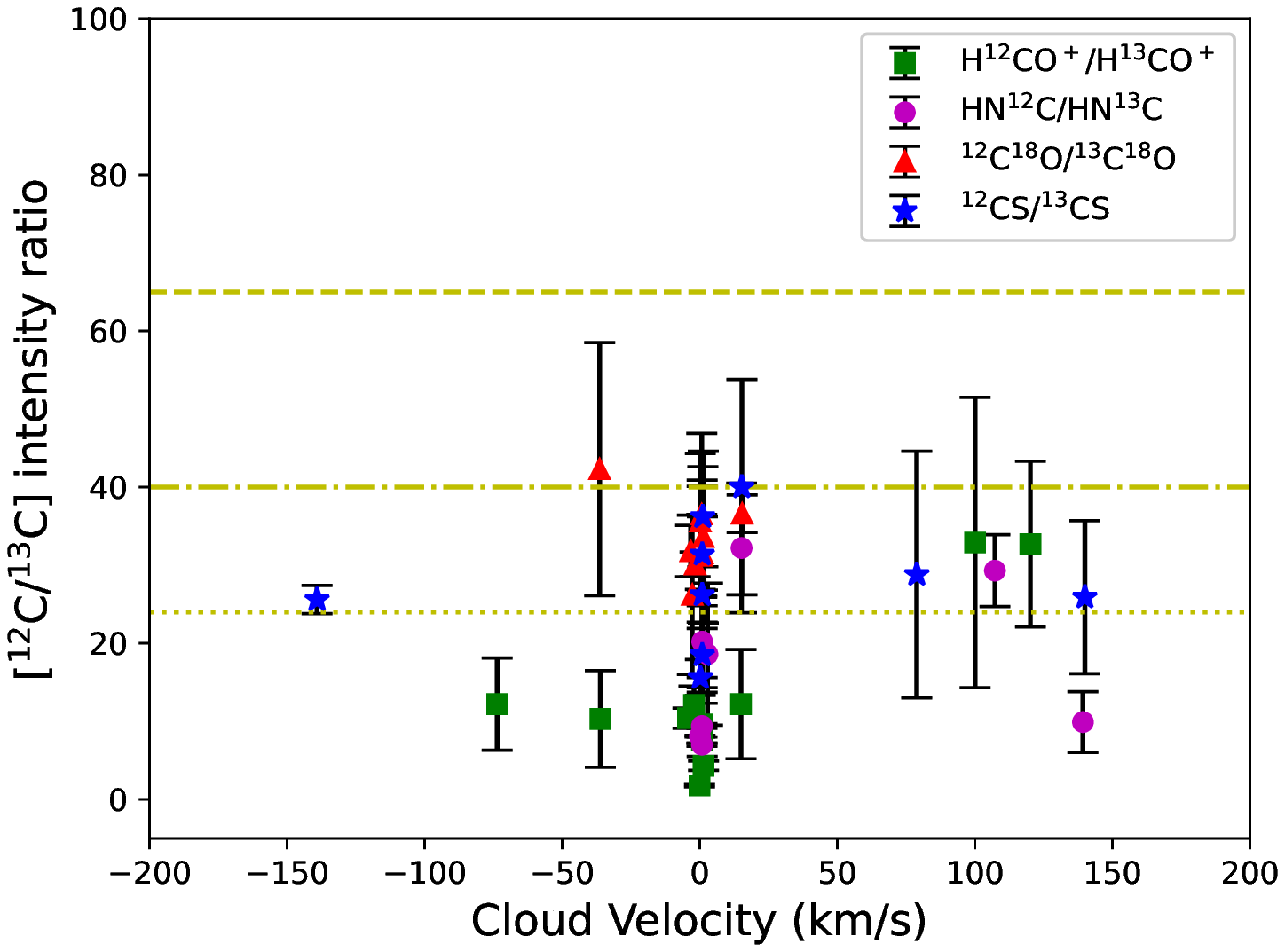} \quad \includegraphics[scale=0.45]{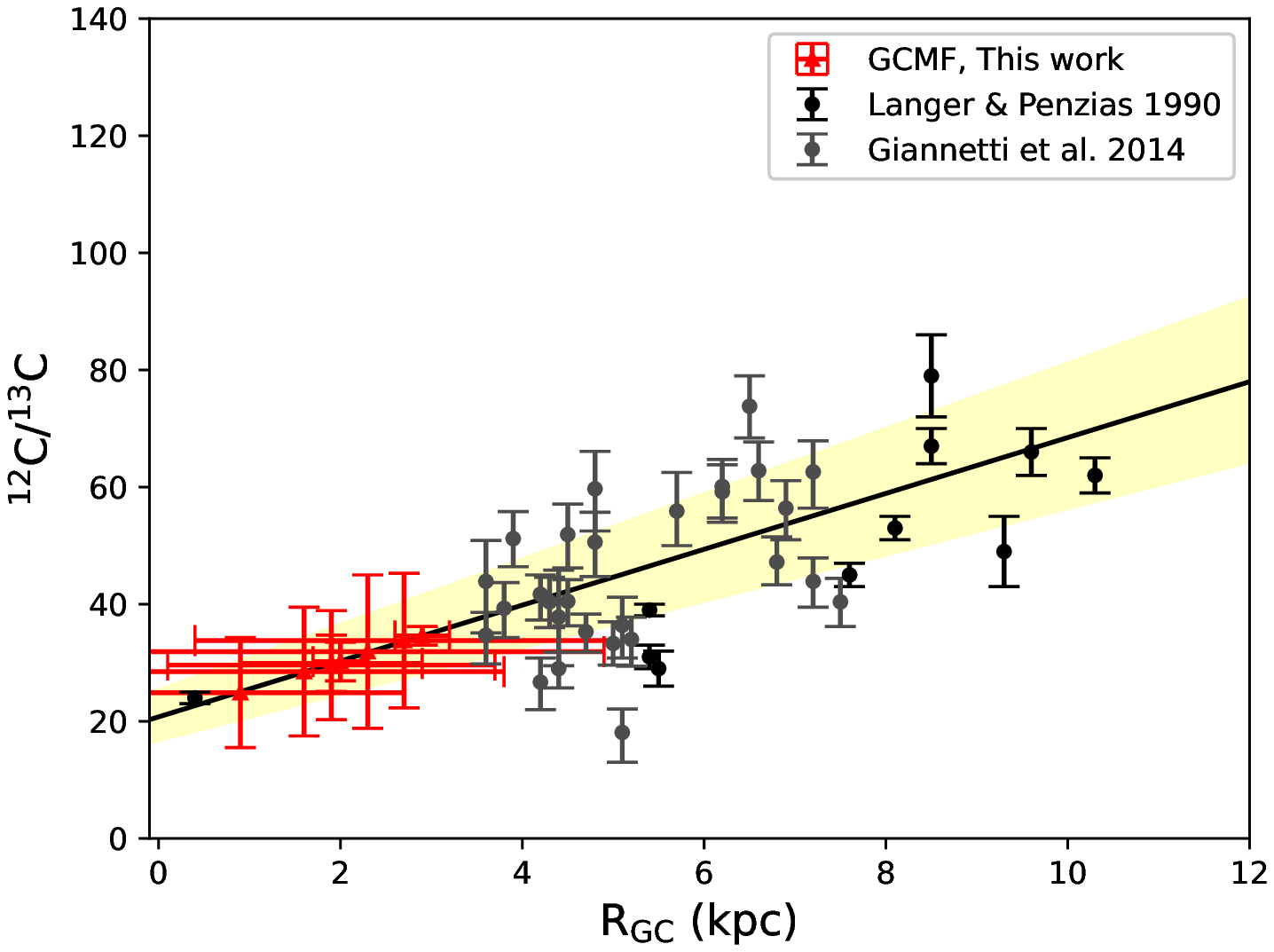} 
\caption{Results of the molecular line analysis of the GCMF.  (Left) $^{12}$C/$^{13}$C ratio of various species versus corresponding cloud velocities. Yellow dotted line corresponds to carbon isotope ratio of 24, that of GC. Dashed-dotted line corresponds to $^{12}$C/$^{13}$C=40 and dashed line corresponds to $^{12}$C/$^{13}$C=65, the carbon isotope ratio of local ISM in the solar neighbourhood. (Right) The $^{12}$C/$^{13}$C ratio derived from I[C$^{18}$O]/I[$^{13}$C$^{18}$O] (for GCMF; red triangles) as a function of Galactocentric distance. Distances to cloud components are determined using the $^{12}$C/$^{13}$C gradient across the Galaxy \citep{2023A&A...670A..98Y}. Shaded area corresponds to the 1$\sigma$ interval of the fit. Grey and black dots correspond to isotope ratios from C$^{18}$O and its isotopologue by \citet{2014A&A...570A..65G} and \citet{1990ApJ...357..477L}, respectively.}
\label{Cisograd}
\end{figure*}

To investigate carbon isotopic ratios in the GCMF, we used intensity ratios of four molecular lines: $^{12}$C$^{18}$O (1--0), $^{12}$CS (2--1), HN$^{12}$C (1--0), H$^{12}$CO$^+$ (1--0) and their $^{13}$C isotopologues. We consider all detections with a signal-to-noise ratio $\gtrsim$ 2.5$\sigma$ where $\sigma$ is the rms noise. Apart from the primary emission line of interest here at $\sim$0 km~s$^{-1}$, high velocity components ($\mid$V$_\textrm{LSR}\mid>100$ km~s$^{-1}$) are also seen toward a few positions. Spectra of $^{12}$CS, $^{12}$C$^{18}$O, H$^{12}$CO$^+$, HN$^{12}$C and their $^{13}$C isotopologues toward few selected positions are shown in Figs.~\ref{cs}, \ref{c18o}, \ref{hcop}, and \ref{hnc}. Table~\ref{moltabcs} lists $^{12}$CS and $^{13}$CS line parameters for selected positions where both lines are detected. Similarly, line parameters of $^{12}$C$^{18}$O and $^{13}$C$^{18}$O lines are listed in Table~\ref{moltabc18o}. Fig.~\ref{Cisograd}(Left) shows the $^{12}$C/$^{13}$C line ratios of different species as a function of their velocities. Considering those components that are only associated with the GCMF (--3.5$\leq$V$_\textrm{LSR}\leq$3.5 km~s$^{-1}$; see Fig.~\ref{Cisogradsub}), the intensity ratios for individual molecules are as follows: $^{12}$C$^{18}$O/$^{13}$C$^{18}$O (26.2--36.5, mean=32.1$\pm$3.1), $^{12}$CS/$^{13}$CS (15.6--36.2, mean=25.6$\pm$3.7), HN$^{12}$C/HN$^{13}$C (7.0--20.2, mean=12.7$\pm$2.0), H$^{12}$CO$^+$/H$^{13}$CO$^+$ (1.8--12.1, mean=7.1$\pm$1.0). 
The HN$^{12}$C/HN$^{13}$C and H$^{12}$CO$^+$/H$^{13}$CO$^+$ ratios are systematically lower than the $^{12}$C$^{18}$O/$^{13}$C$^{18}$O and $^{12}$CS/$^{13}$CS ratios derived at the same positions, and is less than half the standard value at the GC \citep[$\sim$ 25;][and references therein]{2010A&A...523A..51R}. This implies that H$^{12}$CO$^+$ and HN$^{12}$C are mostly optically thick. 

The intensity ratios derived from $^{12}$C$^{18}$O and $^{12}$CS species are significantly lower than the $^{12}$C/$^{13}$C ratio of the local ISM ($\sim$ 65; see Table~\ref{isocsc18o}). The ratios toward P1 and P9 estimated using $^{12}$CS is lower than that of $^{12}$C$^{18}$O ($<20$). Hence $^{12}$CS could be optically thick at these positions whereas ratios are similar to those derived from $^{12}$C$^{18}$O at positions P2 and P6. In general, values derived using $^{12}$C$^{18}$O are the most reliable estimates of carbon isotope ratios as $^{12}$C$^{18}$O is $\sim$500 times less abundant than $^{12}$C$^{16}$O \citep{1997ApJ...482..285B} and hence can safely be assumed to be optically thin. However, the isotope ratio derived from $^{12}$C$^{18}$O and its isotopologue could suffer from fractionation effects. For example, \citet{1984ApJ...277..581L} studied the fractionation effects in dense molecular clouds and found that the carbon isotope ratios can be separated into three groups: (a) CO group, (b) HCO$^+$ group, and (c) the carbon isotope pool group consisting of species such as CS, HCN, etc. They concluded that for group (a), the abundance of $^{13}$CO is enhanced relative to that of $^{12}$CO at lower temperatures, densities and high metal abundances, whereas $^{12}$C is enhanced for species belonging to group (c), the carbon isotope pool. On the other hand, selective photodissociation of $^{13}$CO can increase the $^{12}$C/$^{13}$C ratio in diffuse clouds \citep{2009A&A...503..323V}. Overall, intensity ratios of C$^{18}$O and CS are consistent with each other and thus, we cannot find any significant evidence of these effects in the GCMF. 

\subsection{Distance estimates from carbon isotope ratio}
For determining the distance to the GCMF, we only consider carbon isotope ratios derived from C$^{18}$O/$^{13}$C$^{18}$O. Conversion of the $^{12}$C$^{18}$O/$^{13}$C$^{18}$O intensity ratios to carbon isotopic ratios requires correcting for two factors: 1) difference in frequency between the two isotopes leading to a difference in proportionality constant between column density and antenna temperature \citep{1977ApJ...214...50L} which can be corrected by a multiplicative factor of 0.95 \citep{1981ApJ...243L..47W}, 2) opacity corrections for possible optical depth effects for $^{12}$C$^{18}$O. 
Using J=2--1 and J=1--0 transitions of $^{12}$C$^{18}$O and an LVG radiative transfer model, \citet{1990ApJ...357..477L} estimated opacity corrections toward a sample of interstellar clouds and found a typical correction of 5$\%$. These are dense clouds that host high mass star formation where the optical depth effects could play a significant role compared to the GCMF. If we assume 5$\%$ as the upper limit to the opacity correction, this translates to a maximum isotope ratio uncertainty of $\pm$1.5. To calculate the distance we use the corrected intensity ratios (i.e., 24.9--34.7 with a mean of 30.6$\pm$2.9). We apply the isotope gradient--Galactocentric distance relation derived by \citet{2023A&A...670A..98Y}
\begin{equation}
\mathrm{^{12}C/^{13}C=(4.77\pm0.81)R_{GC}+(20.76\pm4.61)} 
\end{equation}
\noindent where R$_\textrm{GC}$ is the Galactocentric distance. We find R$_\textrm{GC}$ of GCMF in the range 0.9--2.9 kpc (see Fig.~\ref{Cisograd}(Right)) with a mean $\overline{\textrm R}_\textrm{GC}$ 2.0$\pm$0.6 kpc (i.e., heliocentric distance D$=6.2\pm0.6$ kpc). Thus, the current analysis supports the association of the GCMF with the inner 3 kpc of the Galaxy. Using five complex molecules, \citet{2017ApJ...845..158H} estimated $^{12}$C/$^{13}$C ratio toward the GC source SgrB2(N), which was found to range between $15\pm5$ and $33\pm13$, with an average value of $24\pm7$. Their mean estimate, within the uncertainties, is consistent with our estimate of the $^{12}$C/$^{13}$C ratio. Hence, it is likely that the GCMF is located in the GC. However, we would like to caution that the obtained carbon isotope ratios could still suffer from isotopic fractionation effects. Follow up observations are necessary to investigate these effects in detail.
\section{Discussion}
From its funnel-shaped morphology, spatial correlation with the northern lobe of the multiwavelength GC chimney and chemical signatures supporting association with the GC, we believe that GCMF is tracing the molecular outflow from the centre of the Milky Way. The physical extent of the structure, as determined by the heliocentric distance derived from mean isotope ratio (D=6.2 kpc), is estimated to be $\sim103\times108$ pc. If the  GCMF is located at the GC \citep[D=8.2 kpc;][]{Reid2019, 2019A&A...625L..10G}, the physical size of the funnel is $\sim130\times145$ pc. It is important to note that these measurements do not account for potential projection effects, suggesting that the derived sizes serve as lower limits. Overall, the structure appears to extend over 100 pc. Previous studies have identified atomic hydrogen \citep{2013ApJ...770L...4M} and molecular CO counterparts of the outflow from our Galactic nucleus. \citet{2018ApJ...855...33D} discovered a population of anomalous high velocity HI clouds (V$_\textrm{LSR}\leq$ 360 km~s$^{-1}$) extending up to heights of 1.5 kpc ($2.5^\circ <\mid b \mid <10^\circ$) from the Galactic plane whose velocities deviate from purely circular Galactic rotation. These are explained as the relics of dense, cold, neutral gas blown away by the Galactic nuclear wind. CO emission ($+160<$V$_\textrm{LSR}<280$ km~s$^{-1}$) associated with these high velocity HI outflows has also been detected \citep{2020Natur.584..364D}. The $^{13}$CO structure discovered in the SEDIGISM data is at lower latitudes and different velocities ($-3.5<$V$_\textrm{LSR}<3.5$ km~s$^{-1}$), which raises the question of how it is related to the phenomena discussed above.
Using radio recombination lines (RRLs), \citet{2009ApJ...695.1070L} studied the kinematics of the northern GC lobe. They found surprisingly low LSR velocities and velocity widths (V$_\textrm{LSR}$: 0.4--1.7 km~s$^{-1}$, $\Delta$V: 11.5--24.5 km~s$^{-1}$) and proposed that such low velocities could be explained by gas deceleration due to ambient magnetic fields. The CO observations towards the double helix nebula (DHN), believed to be a magnetically shaped feature at the GC (position P6 in Fig.~\ref{13coiso}) also show emission features at 0 km~s$^{-1}$ \citep{2014ApJS..213....8T}. The velocity of GCMF is well in agreement with the RRL velocity measurements of the GC lobe and molecular line velocities of the DHN. Several formation scenarios are possible: For example, the bubble could be (i) advected molecular gas outlining the base of the outflow \citep[e.g.,][]{2015MNRAS.454..238W}, (ii) gas formed via shock compression and/or (iii) material that has cooled down and is either outflowing or falling back onto the Galactic plane \citep{2021MNRAS.506.5658B}. It is possible that the low-velocity, low latitude GCMF is much younger compared to the higher latitude high velocity outflows launched at a different epoch and traces  intermittent activity at the GC. 

From Fig.~\ref{GCbub} and Fig.~\ref{rgb}, it is evident that the GCMF is closely associated with the radio emission from the northern lobe of the 430 pc bipolar radio bubble. Such geometries where CO outflow encasing the ionising gas, is similar to that observed in extragalactic nuclear outflows \citep[e.g.,][]{2023MNRAS.522.3753D}. Moreover, several non thermal radio filaments \citep[NTFs;][]{1984Natur.310..557Y} are located within the cavity of the funnel. This indicates that the progenitor event is also source of the relativistic particles required to power the NTFs. \citet{2019Natur.573..235H} suggested that the observed bipolar radio bubble could be a less energetic version of the event that generated the $\textit{Fermi}$ bubbles. Such events could be powered by past activity from the SgrA* \citep{{2011MNRAS.415L..21Z},{2020ApJ...894..117Z}} and/or supernova driven winds \citep{{2014MNRAS.444L..39L},{2015ApJ...808..107C}}. According to \citet{2015MNRAS.453.3827S}, a star formation rate of $\sim0.5$~M$_\odot$yr$^{-1}$ over a time scale of $\sim$30 Myr could produce $\textit{Fermi}$-like bubbles. This is in agreement with the enhanced star formation rate of 0.2--0.8~M$_\odot$yr$^{-1}$ observed towards the nuclear stellar disc over the past 30~Myrs \citep{2020NatAs...4..377N}. Using 3D magnetohydrodynamic simulations, \citet{2021ApJ...913...68Z} investigated the origin of the 430 pc bipolar radio bubble. A magnetic field of 80 $\mu$G and a supernova rate of 1 kyr$^{-1}$ reproduced the observed morphology, internal energy and X-ray luminosity of the bubbles after an evolutionary time of 330 kyr. This model could also explain a fraction of the NTFs that are believed to originate within mutual collisions between the shock waves of the individual supernovae. \citet{2023arXiv230601071Y} have classified NTFs into two distinct categories: long NTFs with non-thermal spectral indices, oriented perpendicular to the Galactic plane, and short NTFs with thermal spectral indices aligned parallel to the plane. The origin of long NTFs is attributed to a large-scale Galactic wind that created bipolar radio bubbles, which is consistent with the discussed scenario

\begin{figure}
\centering
\includegraphics[scale=0.12]{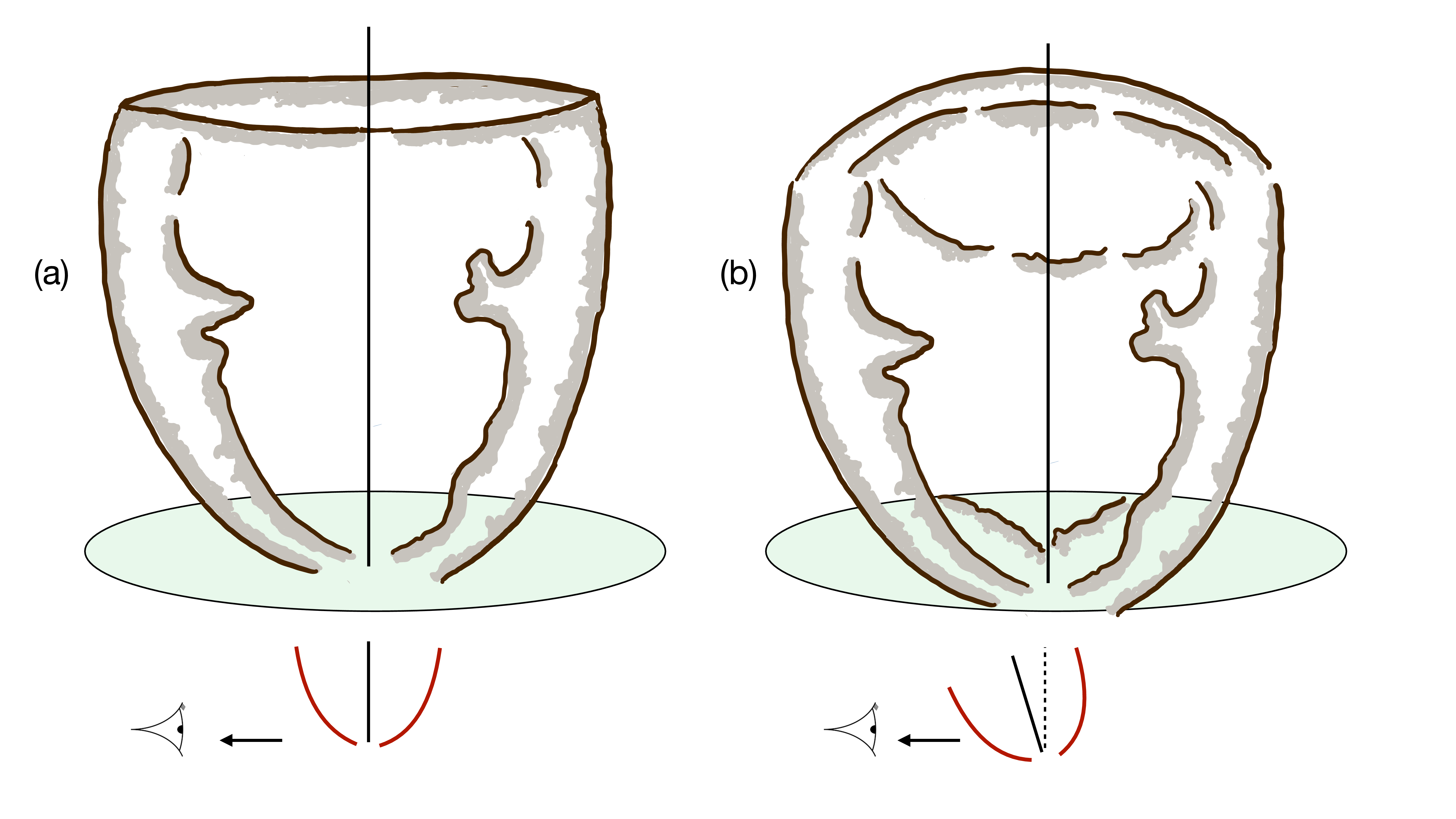} 
\caption{Pictorial representation of the molecular funnel (a) without tilt (b) tilted towards the observer's line of sight so that the base of the funnel appears to be lower than its true position.}
\label{shell}
\end{figure}

Assuming that the funnel-shaped morphology is due to gas expansion out of the plane, we expect the base of the funnel is anchored in the Galactic mid-plane. In this scenario, the narrowest parts of the funnel are high density regions in the Galactic plane and the widest parts are regions of low-density gas in free expansion. The base of the GCMF is located at $b=-0.16^\circ$ (Fig~.\ref{rgb}). Thus at a distance of 8.2 kpc, it is located at 16 pc below SgrA$^*$ (see Fig.~\ref{13coiso}). There are two possibilities for the observed shift with respect to Sgr A$^*$: (a) The funnel is tilted towards our line of sight so that the base appears at  a lower latitude than its true position without the tilt (see Fig.~\ref{shell}); (b) the funnel is located in front of the GC. The IAU mid-plane is tilted ($\theta_\textrm{tilt}$=0.12$^\circ$) with respect to the true mid-plane \citep{2014ApJ...797...53G}. Correcting for the Sun's height above the plane (z$_\textrm{Sun} \sim$25 pc) and tilt of the IAU plane, we find that true mid-plane is located at a distance of D$\sim5.2$ kpc for b=--0.16$^\circ$. This would suggest that the GCMF is located at R$_\textrm{GC}$=3 kpc. A combination of both scenarios is possible as well. However, the possibility of the GCMF being a local cloud with the current assumptions is unlikely as at a heliocentric distance of 2 kpc, the base of the GCMF should appear at $b=-0.6^\circ$. The northern chimney is observed as a silhouette against the low frequency radio emission. Based on this, \citet{2021A&A...646A..66P} suggested that the chimney is rooted at the GC, but is significantly tilted towards us which is in agreement with scenario (a). To summarise, our analysis favours GCMF to be located within the inner 3 kpc of the GC. With the data presented in this paper, we are currently unable to confirm whether the outflow has a southern counterpart as the CO emission is contaminated by emission from foreground star forming regions at negative Galactic latitudes whereas the northern lobe experiences minimal contamination  as it extended toward positive latitudes. Galactic outflows tend to be bipolar, and we are carrying out further investigations to find signatures of a southern lobe.
\section{Conclusions}
We identify a $136\times144$ pc$^2$ $^{13}$CO(2--1) Galactic Centre Molecular Funnel (GCMF). The GCMF extends up to 150 pc above the Galactic plane and is asymmetric with respect to SgrA$^*$, shifted to negative longitudes. The velocity of the gas in the GCMF ranges from $-$3.5 to 3.5 km~s$^{-1}$ and is consistent with the velocity of the ionised gas associated with the northern GC lobe estimated using RRLs. The $^{12}$C/$^{13}$C isotope ratios (24.9--34.7; mean=$30.6\pm2.9$) is consistent with the structure located at the GC or within inner 3 kpc of the Galaxy. The GCMF appears to encompass radio and MIR emission corresponding to the northern lobe of the GC chimney and has finger-like protrusions radially pointed inward indicating their origin is related to a GC outflow in which molecular gas can emerge via dense gas advection, shock compression, cooling flows, or a combination of them. Combining our observational results with previous studies, we propose that the GCMF is the molecular counterpart of the northern Galactic chimney. This gives us a unique opportunity to  study the properties of molecular gas associated with GC chimney which can also be used as a local template to understand the chemistry of nuclear outflows in external galaxies.

\begin{acknowledgements}
We thank the referee for the valuable comments and suggestions that improved the quality of the paper. We thank the staff of IRAM 30m telescope for their excellent support.  This publication is based on data acquired with the Atacama Pathfinder Experiment (APEX) under programmes 092.F-9315 and 193.C-0584. APEX is a collaboration among the Max-Planck-Institut fur Radioastronomie, the European Southern Observatory, and the Onsala Space Observatory. The processed data products are available from the SEDIGISM survey database located at https://sedigism.mpifr-bonn.mpg.de/index.html, which was constructed by James Urquhart and hosted by the Max Planck Institute for Radio Astronomy. This work is based in part on observations made with the Spitzer Space Telescope, which was operated by the Jet Propulsion Laboratory, California Institute of Technology under a contract with NASA. This work make use of data from MeerKAT telescope. The MeerKAT telescope is operated by the South African Radio Astronomy Observatory, which is a facility of the National Research Foundation, an agency of the Department of Science and Innovation.     
\end{acknowledgements}

\bibliographystyle{aa}
\bibliography{astro}
\clearpage
\newpage
\onecolumn
\appendix
\section{Additional Figures and Tables}
\begin{figure*}[h!]
\centering
\includegraphics[scale=0.87]{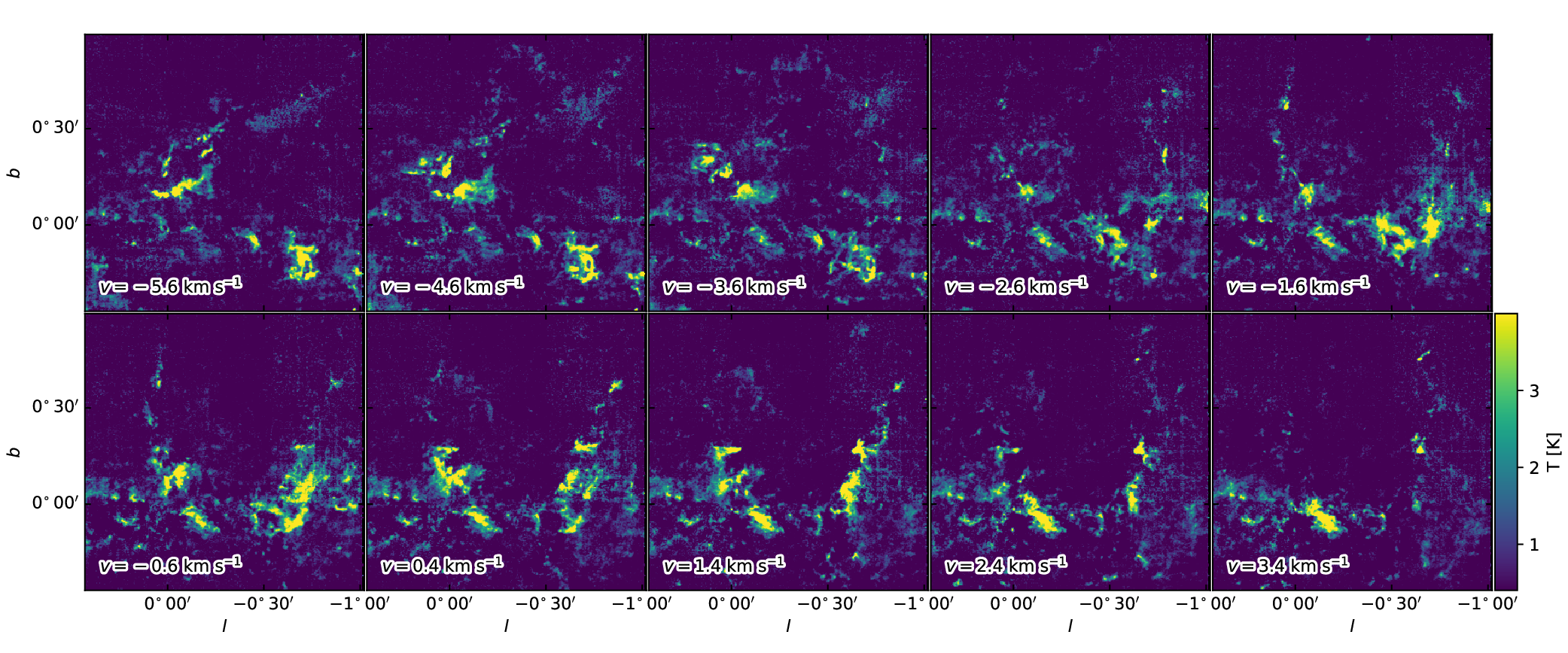}  \quad  
\caption{SEDIGISM $^{13}$CO channel maps in the velocity range --5.6 to 3.4 km~s$^{-1}$ showing the funnel shaped molecular feature in the GC.}
\label{channel}
\end{figure*}
\begin{figure*}[h!]
\centering
\includegraphics[scale=0.35]{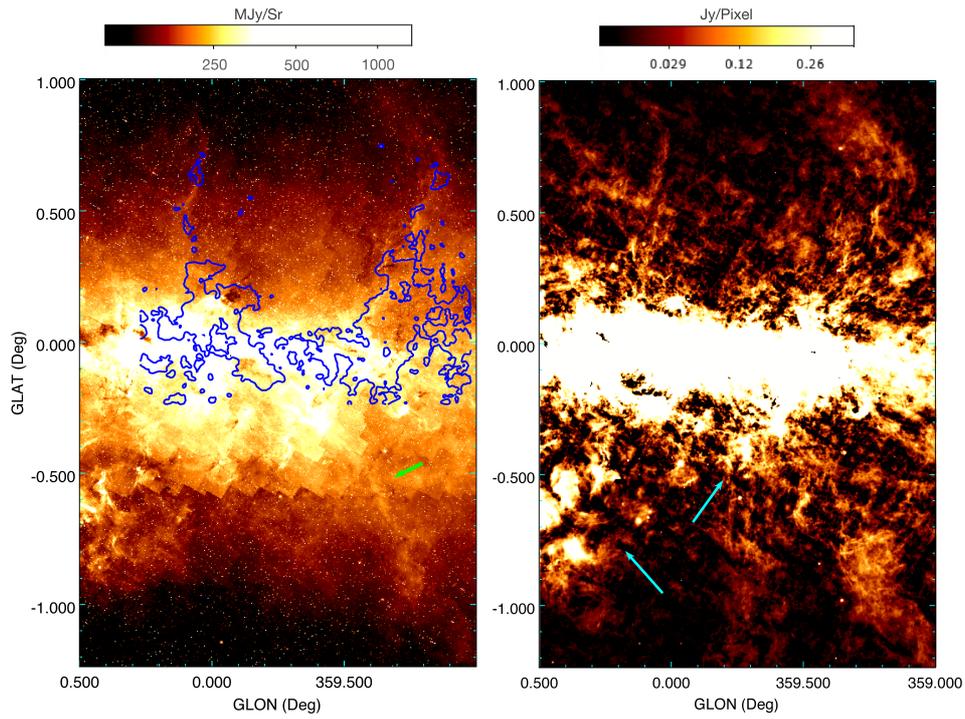} 
\caption{(Left) Spitzer 8~$\mu$m emission overlaid with $^{13}$CO contours. The contour level is 4 K\,kms$^{-1}$. Green arrow points to part of the southern lobe of the bipolar chimney. (Right) Herschel 70~$\mu$m emission. Cyan arrows point to foreground \hii~regions.}
\label{funnelinfra}
\end{figure*}
\begin{figure*}[h!]
\centering
\hspace*{-0.6 cm}
\includegraphics[scale=0.16]{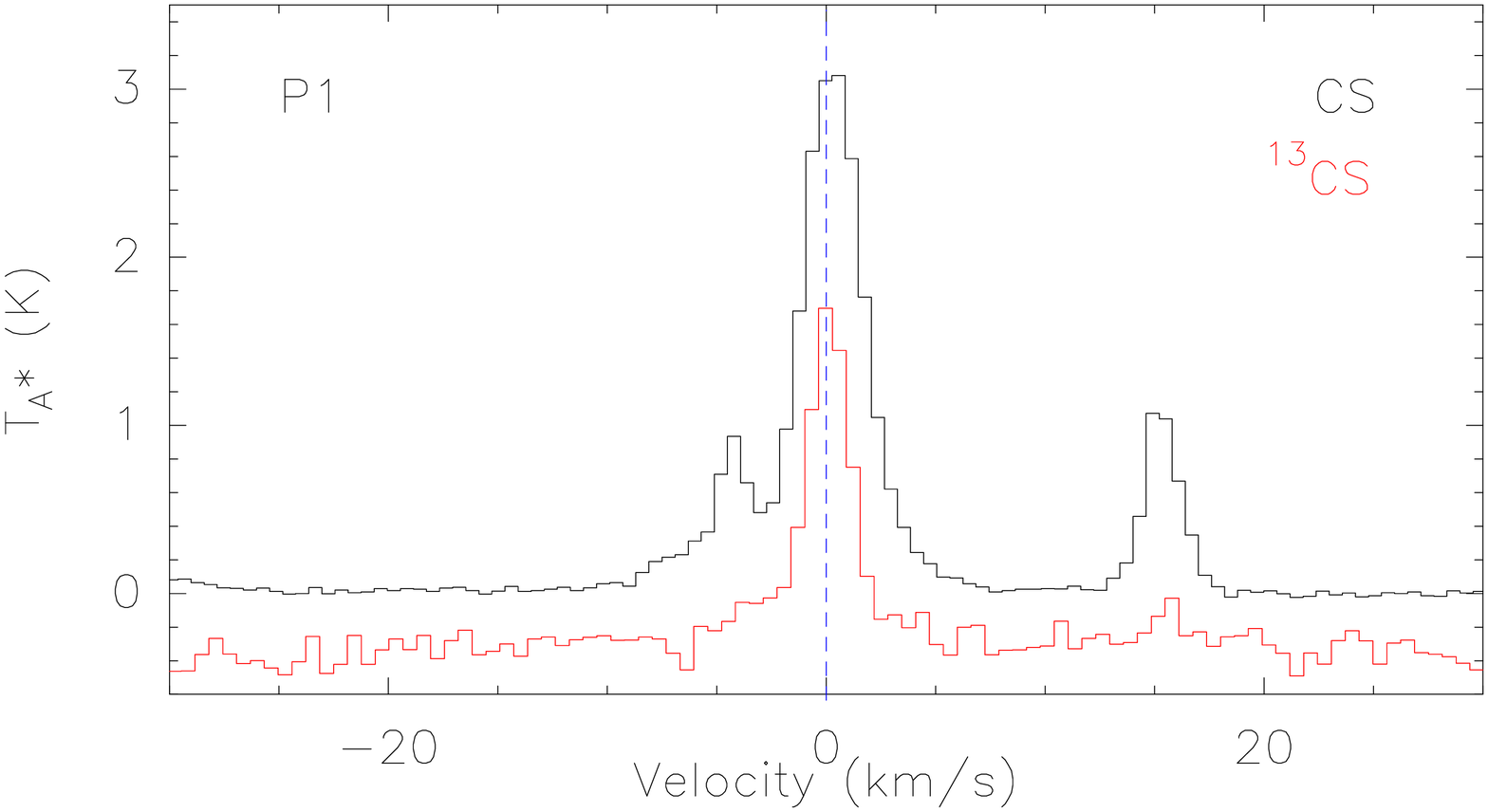}\quad \includegraphics[scale=0.16]{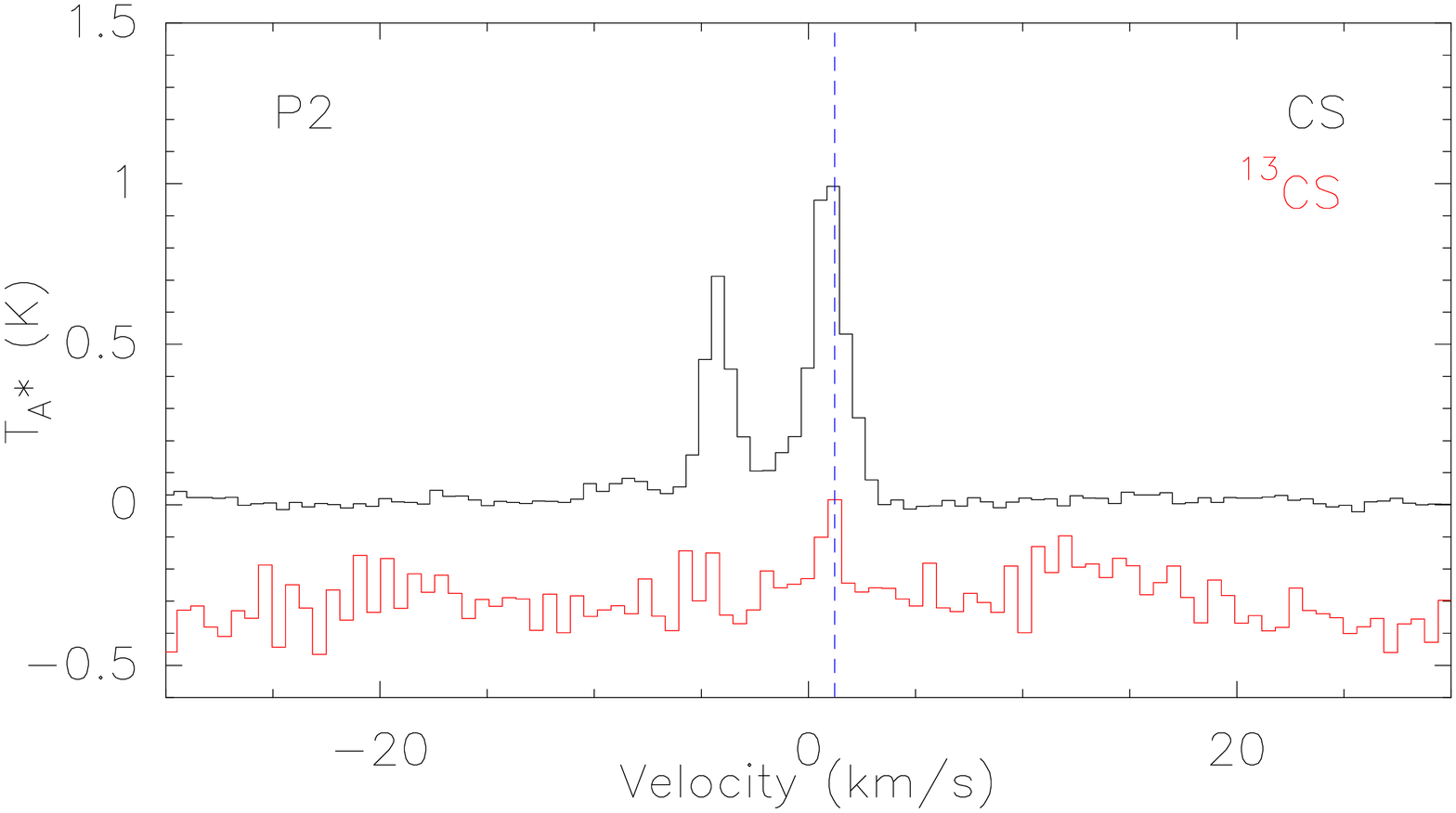} \quad \includegraphics[scale=0.16]{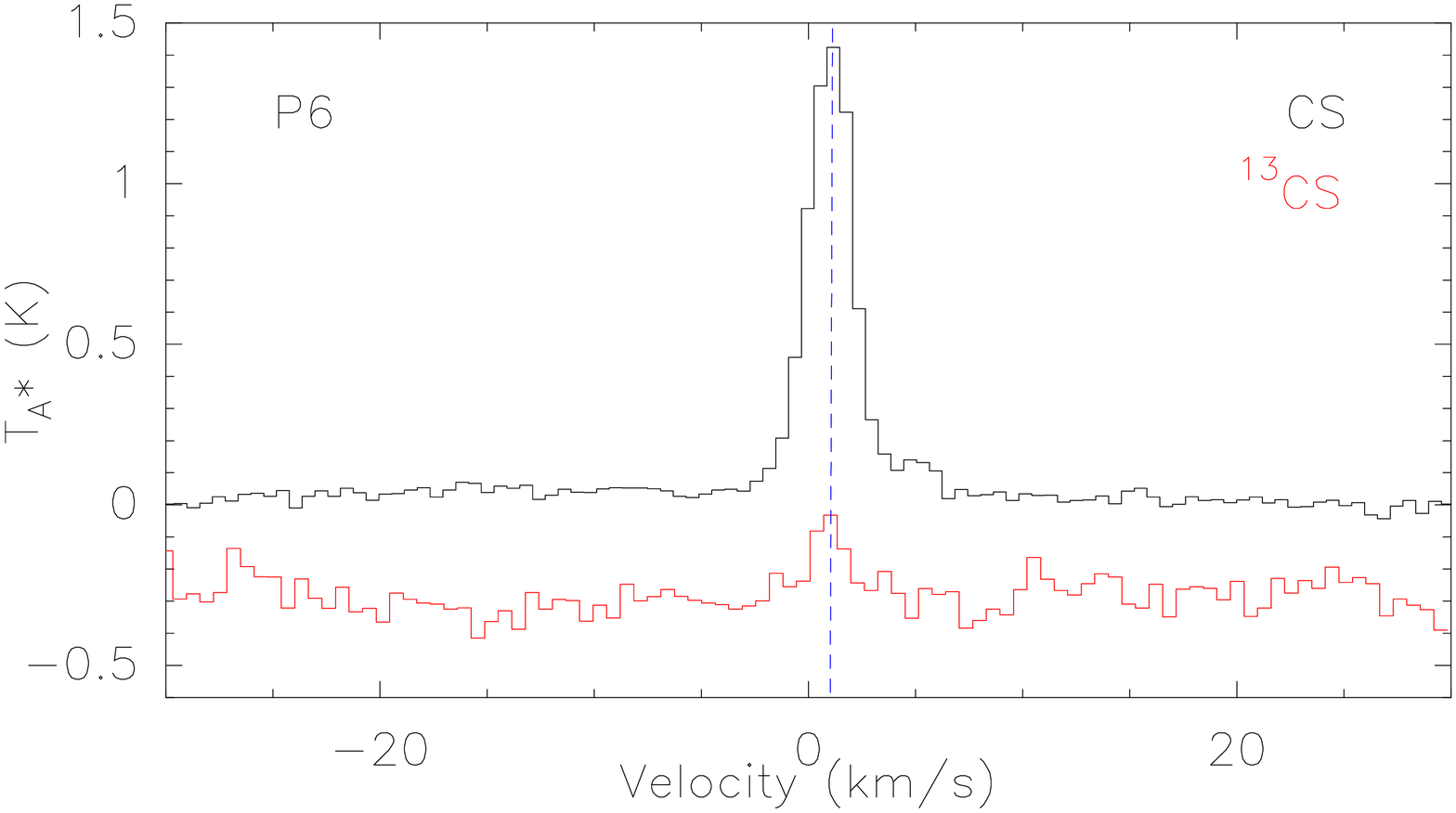} \quad \includegraphics[scale=0.16]{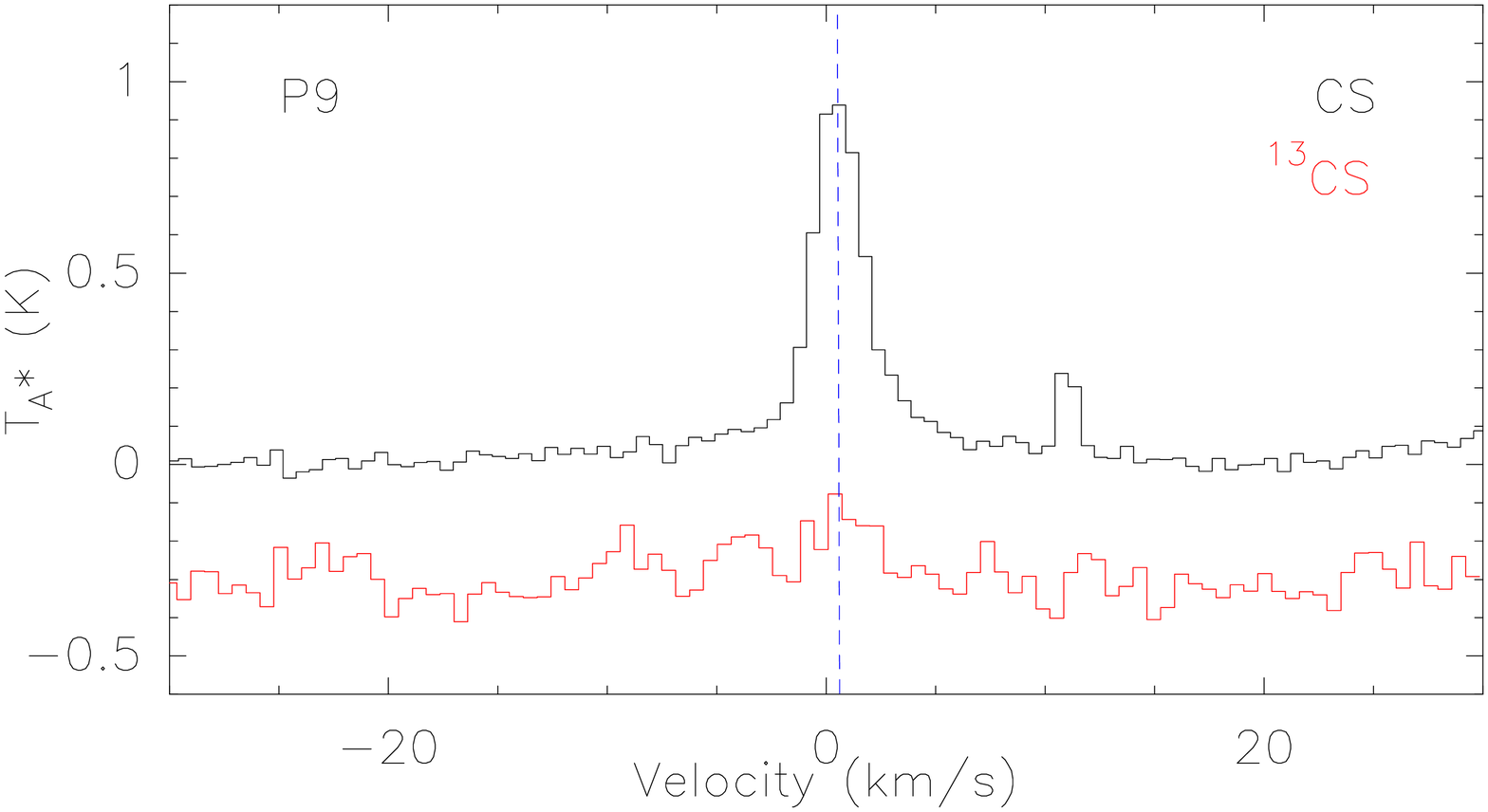}
\caption{Spectra of CS and $^{13}$CS $J=2-1$ emission toward 4 positions along the GCMF restricted to a velocity range of $-30$ to 30 km~s$^{-1}$ for a better visualisation of the emission features. $^{13}$CS spectra are scaled by a factor of 6. \textbf{Blue dashed line corresponds to peak of $^{13}$CS line.}}
\label{cs}
\end{figure*}
\begin{figure*}[]
\centering
\includegraphics[scale=0.16]{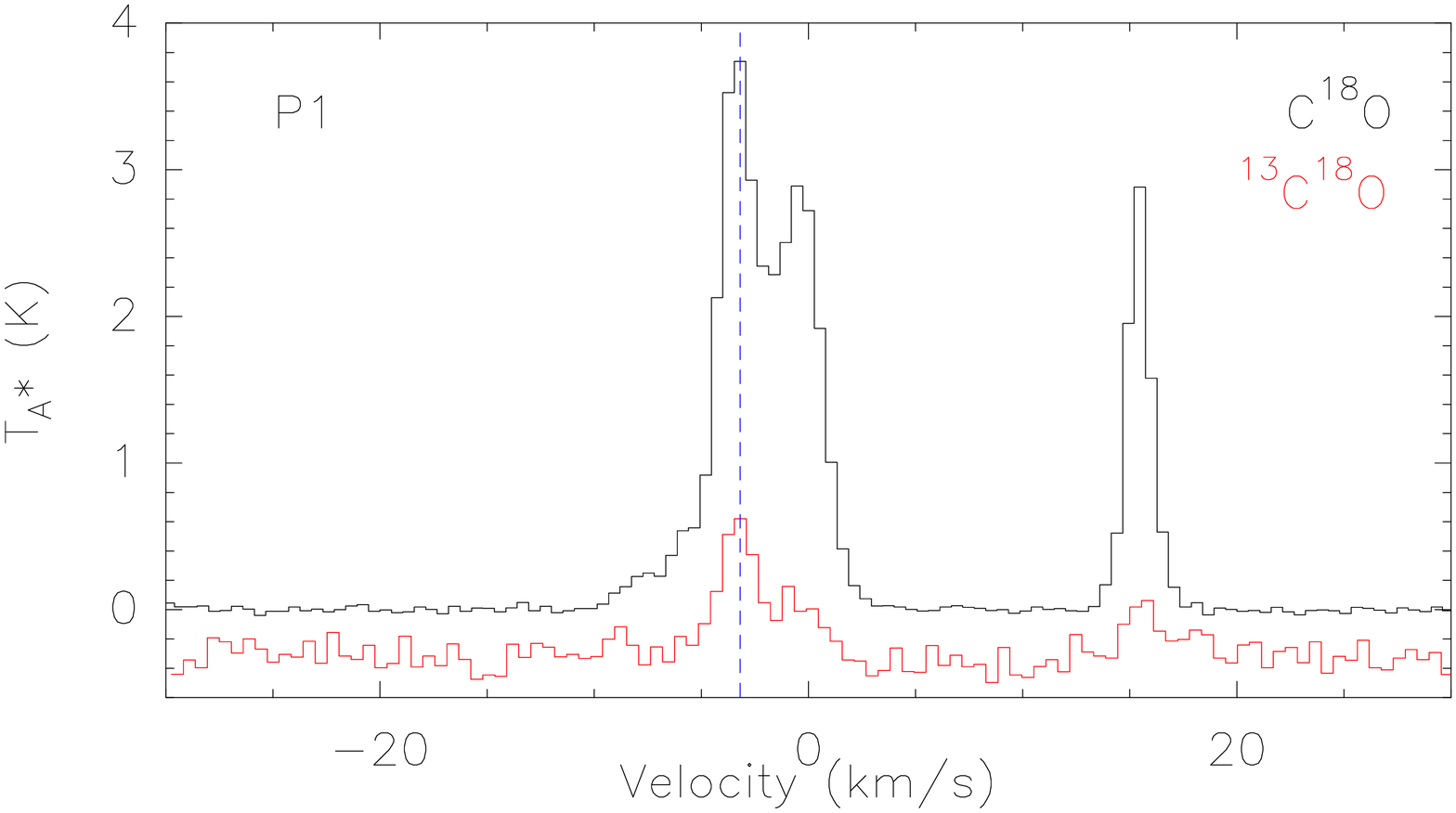}\quad \includegraphics[scale=0.16]{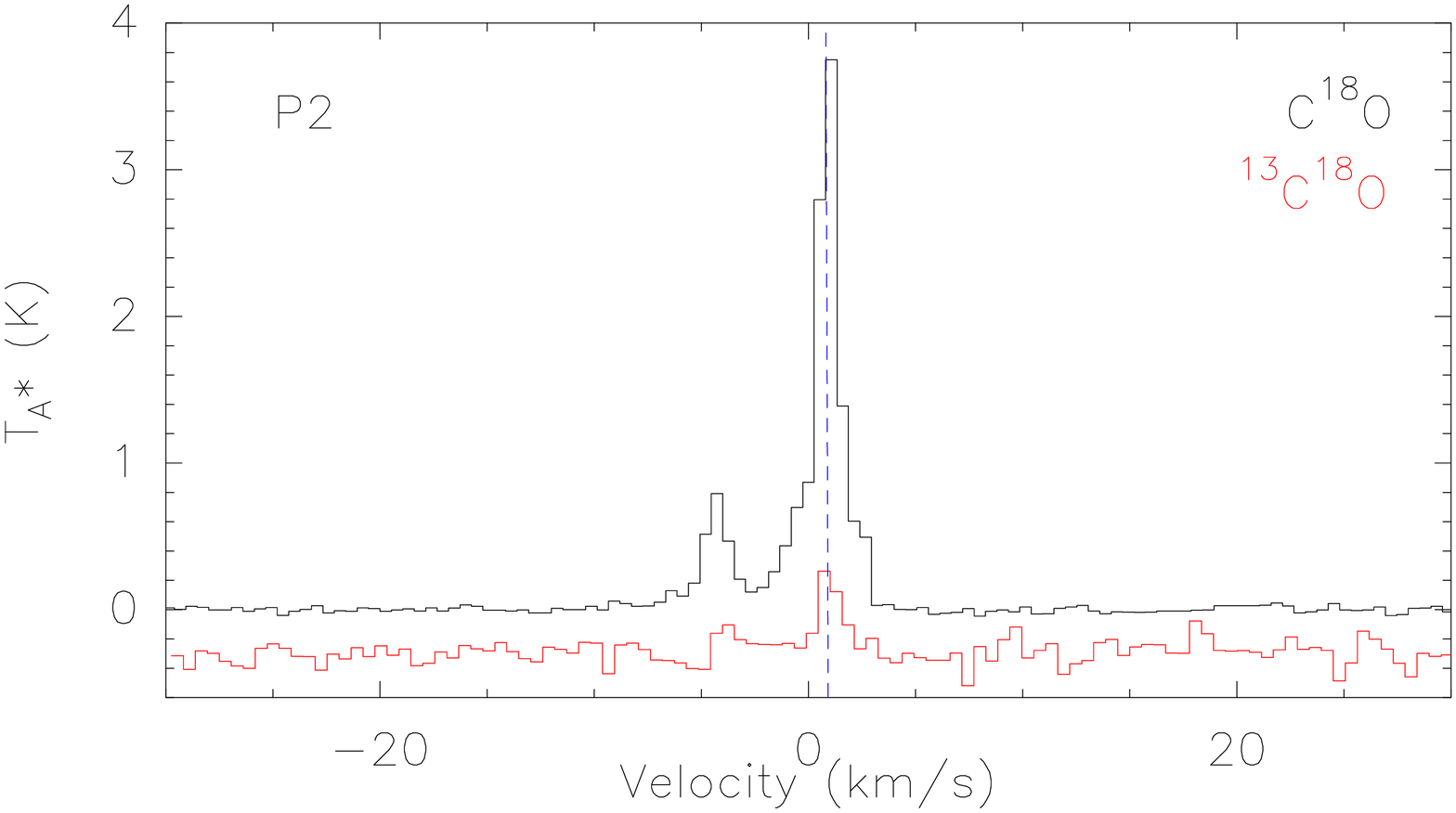} \quad \includegraphics[scale=0.16]{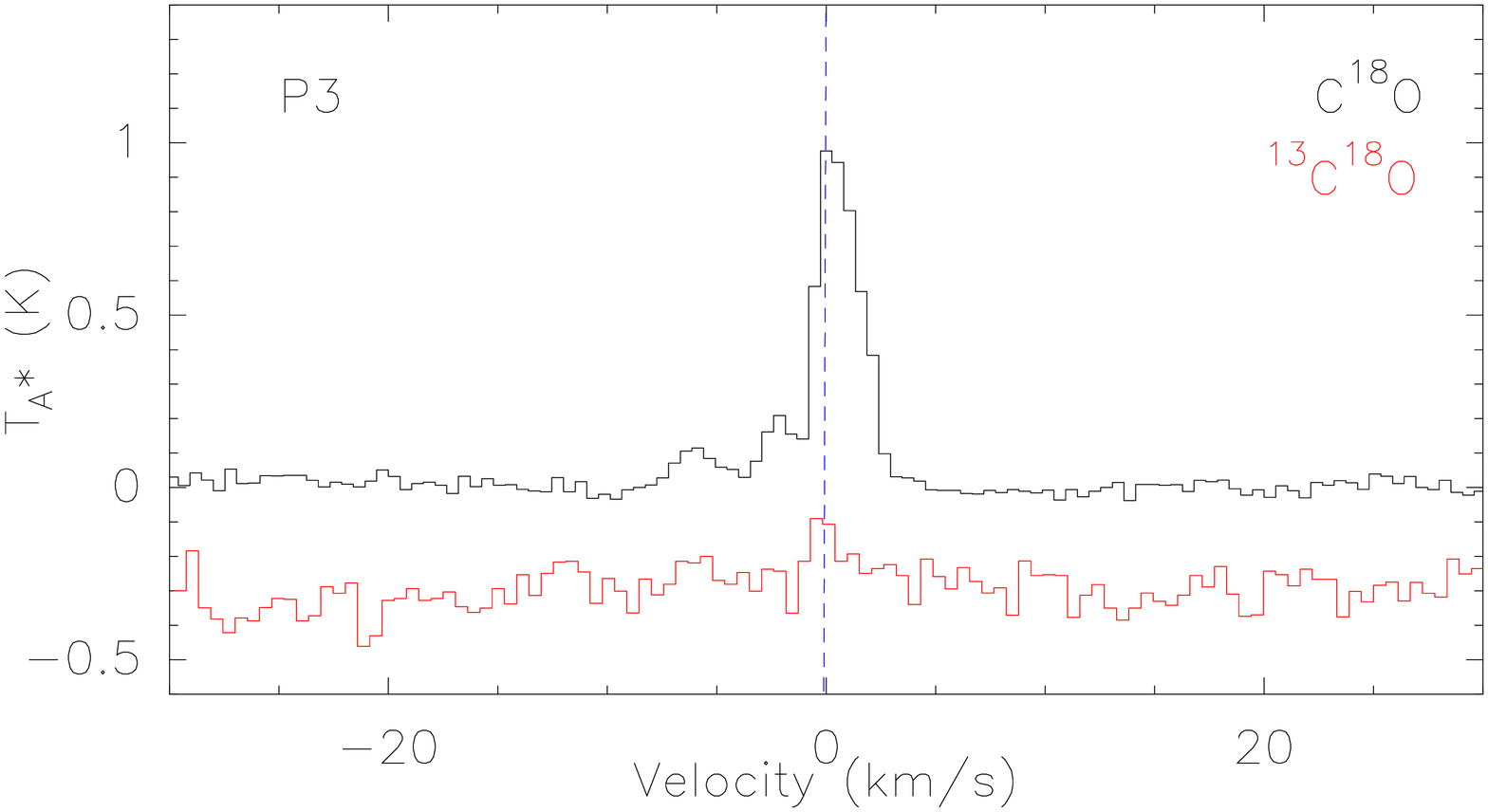} \quad \includegraphics[scale=0.16]{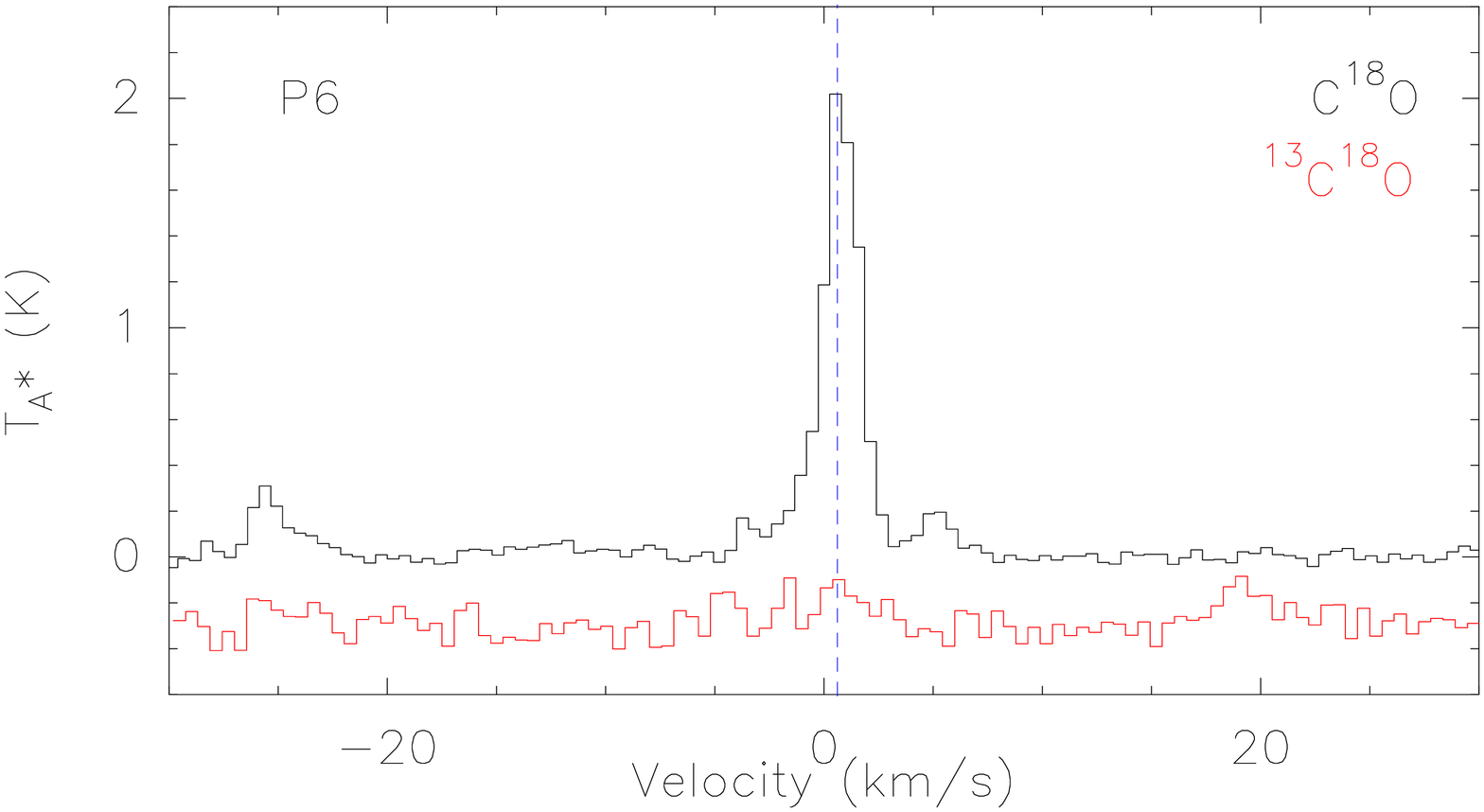} \quad \includegraphics[scale=0.16]{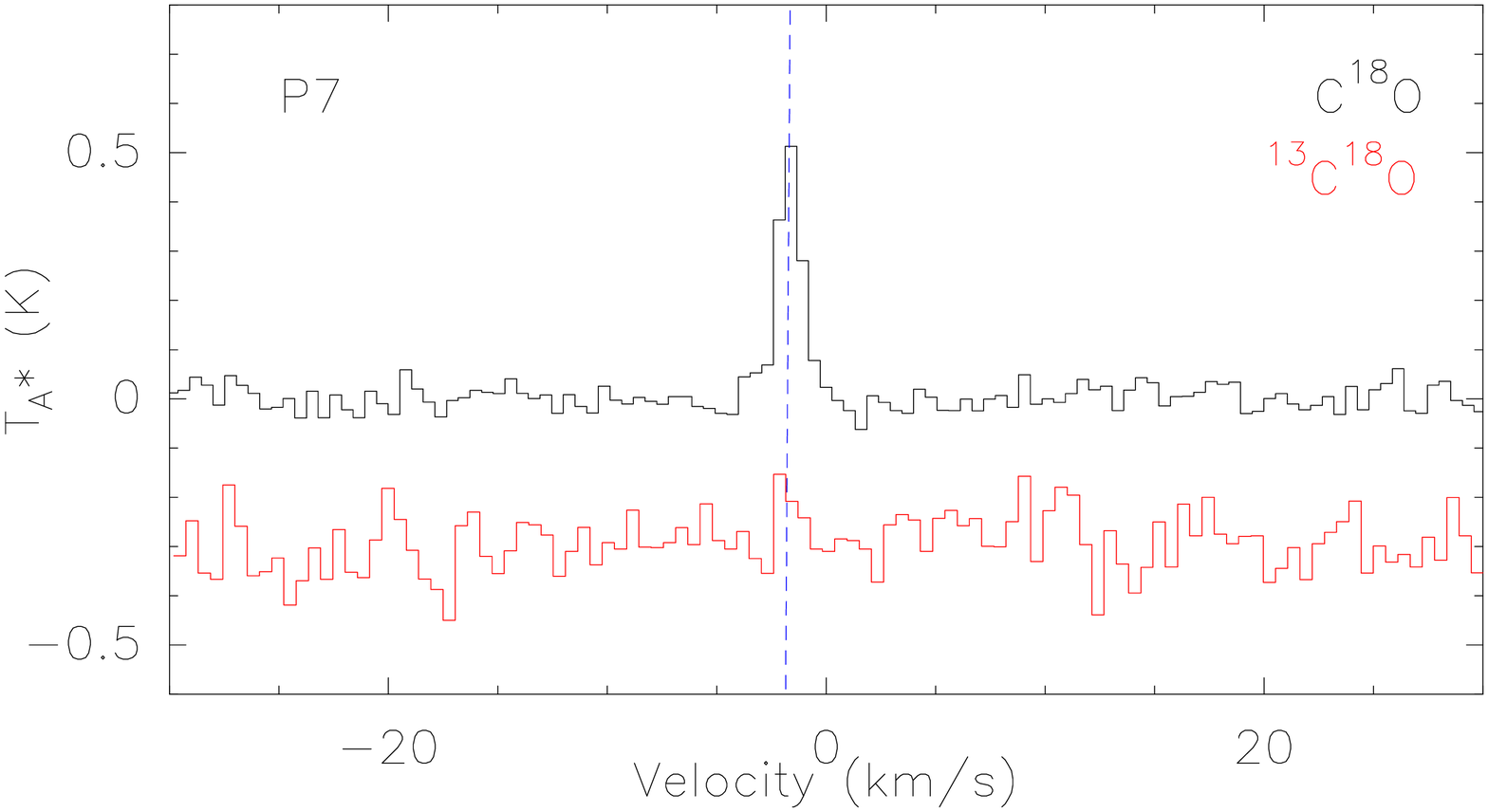} \quad 
\includegraphics[scale=0.16]{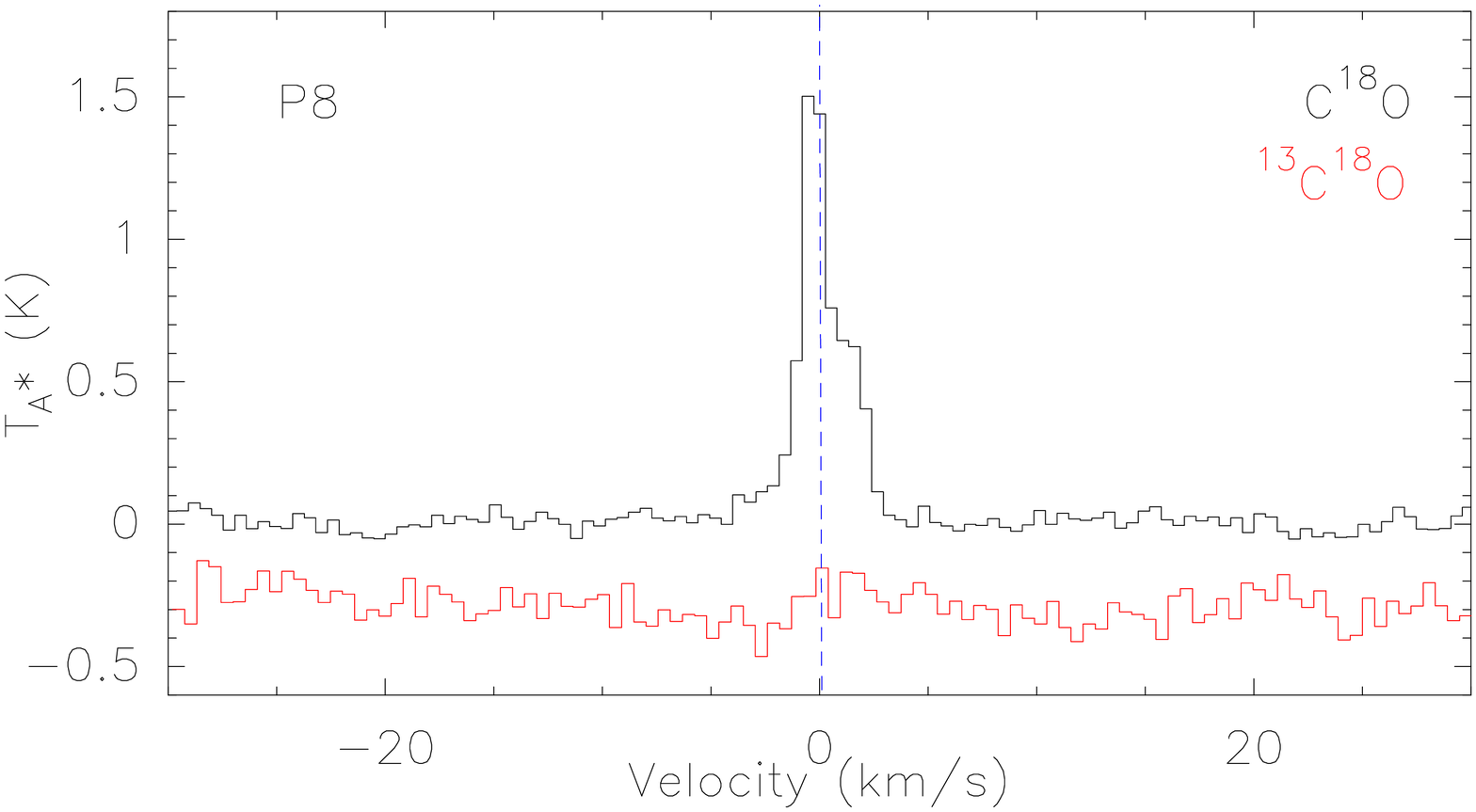} \quad \includegraphics[scale=0.16]{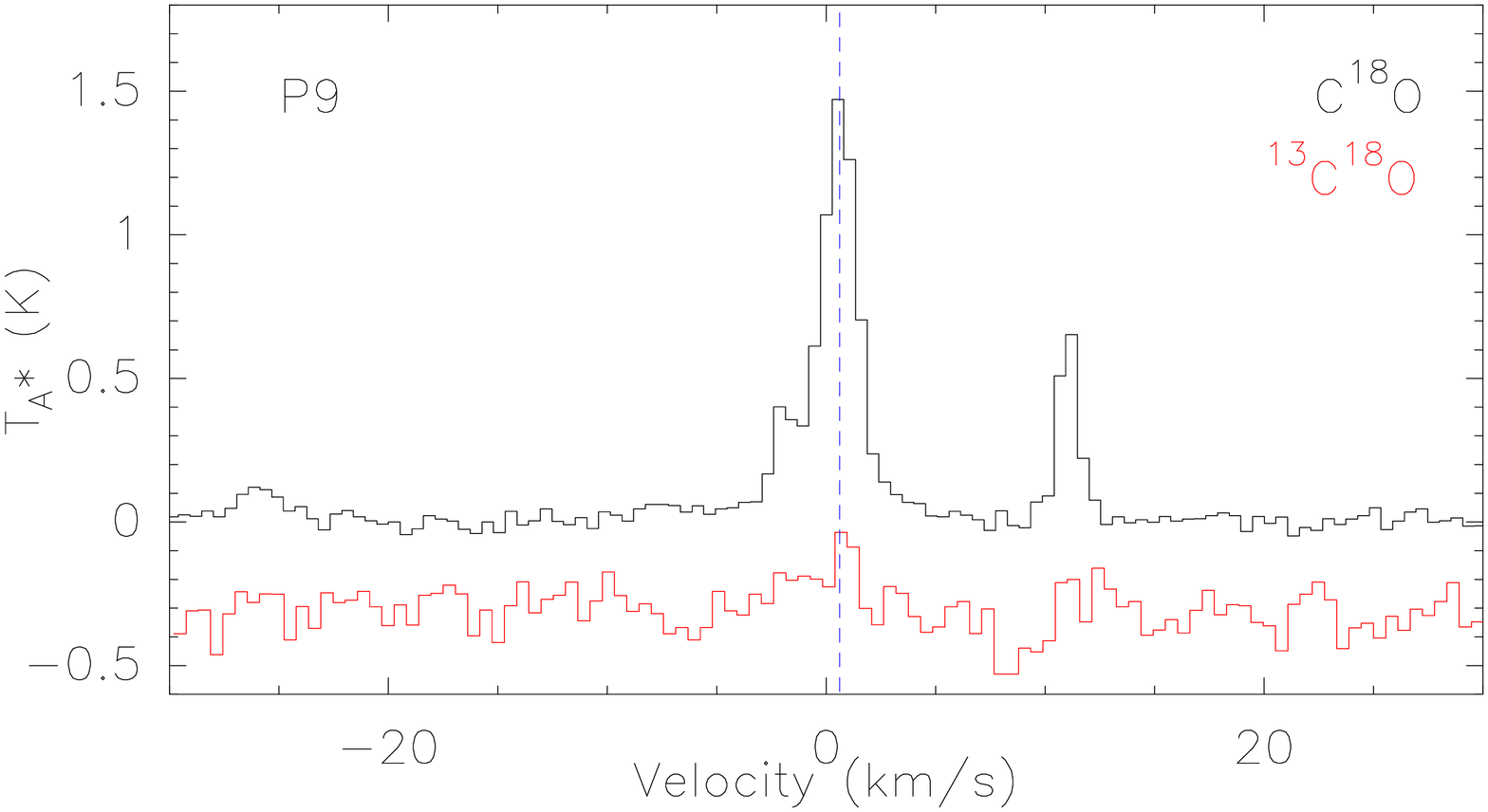} \quad \includegraphics[scale=0.16]{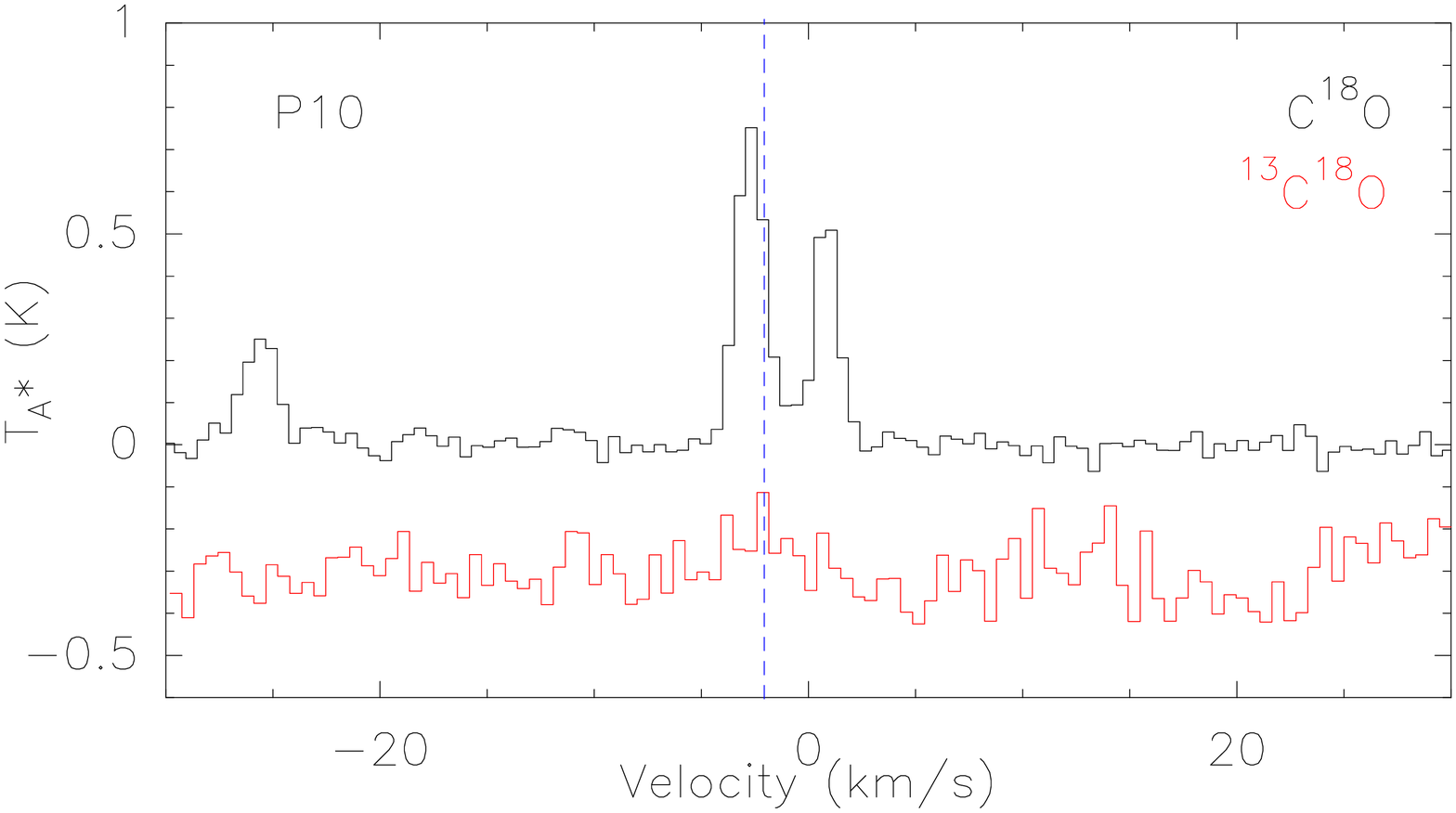}
\caption{Spectra of C$^{18}$O and $^{13}$C$^{18}$O $J=1-0$ emission toward 8 positions along the GCMF restricted to a velocity range of $-30$ to 30 km~s$^{-1}$ for a better visualisation of the emission features. $^{13}$C$^{18}$O spectra are scaled by a factor of 6. \textbf{Blue dashed line corresponds to peak of $^{13}$C$^{18}$O line.}}
\label{c18o}
\end{figure*}
\begin{table*}[h!]
\centering
\caption{Line parameters for CS and $^{13}$CS$^*$.}
    \begin{tabular}{c c c c c c c c c}
 \hline
    \hline\\
Position&&CS&&&&$^{13}$CS&&\\
\hline\\
&$\int \textrm T_A^*\,\textrm{dV}$&T$_\textrm A^*$ & V$_\textrm{LSR}$& $\Delta \textrm V$&$\int \textrm T_A^*\,\textrm{dV}$&T$_\textrm A^*$ & V$_\textrm{LSR}$& $\Delta \textrm V$ \\
&(K\,km~s$^{-1}$)&(K)&(km~s$^{-1}$)& (km~s$^{-1}$)&(K\,km~s$^{-1}$)&(K)&(km~s$^{-1}$)& (km~s$^{-1}$)\\
\hline \\
P1&9.18$\pm$1.03&2.86&$0.24\pm$0.03&3.01$\pm$0.10&0.59$\pm$0.06&0.28&$0.16\pm$0.04&$1.95\pm0.12$\\
&2.32$\pm$0.01&1.10&$15.32\pm$0.01&1.99$\pm$0.05&0.06$\pm$0.02&0.05&$15.62\pm$0.19&$1.19\pm0.64$\\
&6.99$\pm$1.21&0.46&$78.88\pm$1.19&14.20$\pm$2.89&0.24$\pm$0.12&0.02&$79.07\pm$2.95&$10.41\pm8.66$\\
P2&2.09$\pm$0.06&1.00&$0.92\pm$0.03&1.96$\pm$0.07&0.07$\pm$0.02&0.06&$1.04\pm$0.15&$1.14\pm0.49$\\
P6&3.02$\pm$0.02&1.24&$1.07\pm$0.01&2.29$\pm$0.04&0.08$\pm$0.01&0.05&$0.94\pm$0.15&$1.71\pm0.37$\\
P9&1.92$\pm$0.17&0.11&$0.93\pm$0.65&15.99$\pm$2.33&0.10$\pm$0.02&0.03&$0.84\pm 0.29$&$2.96\pm0.75$\\
&15.18$\pm$0.47&0.79&$-139.10\pm$0.27&18.04$\pm$0.67&0.59$\pm$0.04&0.04&$-138.62\pm$0.45&$13.30\pm0.86$\\
&2.12$\pm$0.42&0.19&$140.04\pm$0.82&10.16$\pm$2.88&0.08$\pm$0.03&0.01&$146.79\pm$1.22&$6.69\pm2.25$\\
\hline\\
\multicolumn{9}{l}{\tiny{$^*$listed parameters correspond to positions where both CS and $^{13}$CS have been detected simultaneously.}} 
\end{tabular}
\label{moltabcs}
\end{table*}
\begin{table*}[h!]
\centering
\caption{Line parameters for C$^{18}$O and $^{13}$C$^{18}$O$^*$. }
    \begin{tabular}{c c c c c c c c c}
 \hline
    \hline\\
Position&&C$^{18}$O&&&&$^{13}$C$^{18}$O&&\\
\hline\\
&$\int \textrm T_A^*\,\textrm{dV}$&T$_\textrm A^*$ & V$_\textrm{LSR}$& $\Delta \textrm V$&$\int \textrm T_A^*\,\textrm{dV}$&T$_\textrm A^*$ & V$_\textrm{LSR}$& $\Delta \textrm V$ \\
&(K\,km~s$^{-1}$)&(K)&(km~s$^{-1}$)& (km~s$^{-1}$)&(K\,km~s$^{-1}$)&(K)&(km~s$^{-1}$)& (km~s$^{-1}$)\\
\hline \\
P1&7.12$\pm$0.06&3.43&--3.28$\pm$0.01&1.95$\pm$0.02&0.33$\pm$0.02&0.15&$-3.27\pm$0.07&$1.92\pm0.28$\\
&7.65$\pm$0.05&2.88&--0.44$\pm$0.01&2.50$\pm$0.02&0.05$\pm$0.04&0.06&$-0.99\pm$0.28&$0.7\pm0.48$\\
&0.08$\pm$0.02&0.01&0.65$\pm$0.09&14.79$\pm$8.9&0.08$\pm$0.02&0.05&$0.09\pm$0.55&$1.56\pm1.02$\\

&4.03$\pm$0.03&2.83&15.41$\pm$0.01&1.34$\pm$0.01&0.11$\pm$0.02&0.06&$15.61\pm$0.16&$1.63\pm0.42$\\
P2&4.09$\pm$0.18&3.69&$0.96\pm$0.04&1.05$\pm$0.02&0.13$\pm$0.02&0.10&$0.98\pm$0.09&$1.25\pm0.22$\\
&1.69$\pm$0.24&1.24&$-36.39\pm$0.53&1.29$\pm$0.53&0.04$\pm$0.01&0.06&$-36.36\pm$0.08&$0.60\pm0.94$\\
P3&1.75$\pm$0.05&1.00&$0.24\pm$0.01&1.64$\pm$0.05&0.05$\pm$0.01&0.04&$-0.21\pm$0.16&$1.16\pm0.34$\\
P6&2.56$\pm$0.01&0.98&$0.73\pm$0.02&2.43$\pm$0.04&0.07$\pm$0.02&0.03&$0.86\pm$0.33&$2.28\pm0.96$\\
P7&0.71$\pm$0.03&0.52&$-1.66\pm$0.03&1.29$\pm$0.07&0.02$\pm$0.01&0.04&$-1.89\pm$0.08&$0.56\pm0.20$\\
P8&1.05$\pm$0.07&0.56&$1.33\pm$0.01&1.75$\pm$0.12&0.03$\pm$0.01&0.03&$1.67\pm$0.27&$1.16\pm0.51$\\
P9&1.87$\pm$0.06&1.14&$0.69\pm$0.02&1.54$\pm$0.04&0.06$\pm$0.02&0.05&$0.85\pm 0.15$&$1.06\pm0.381$\\
P10&1.31$\pm$0.05&0.75&$-2.69\pm$0.03&1.63$\pm$0.08&0.05$\pm$0.02&0.02&$-2.58\pm$0.47&$2.20\pm0.99$\\
\hline\\
\multicolumn{9}{l}{\tiny{$^*$listed parameters correspond to positions where both CS and $^{13}$CS have been detected simultaneously.}} 
\end{tabular}
\label{moltabc18o}
\end{table*}
\begin{figure*}[]
\centering
\includegraphics[scale=0.16]{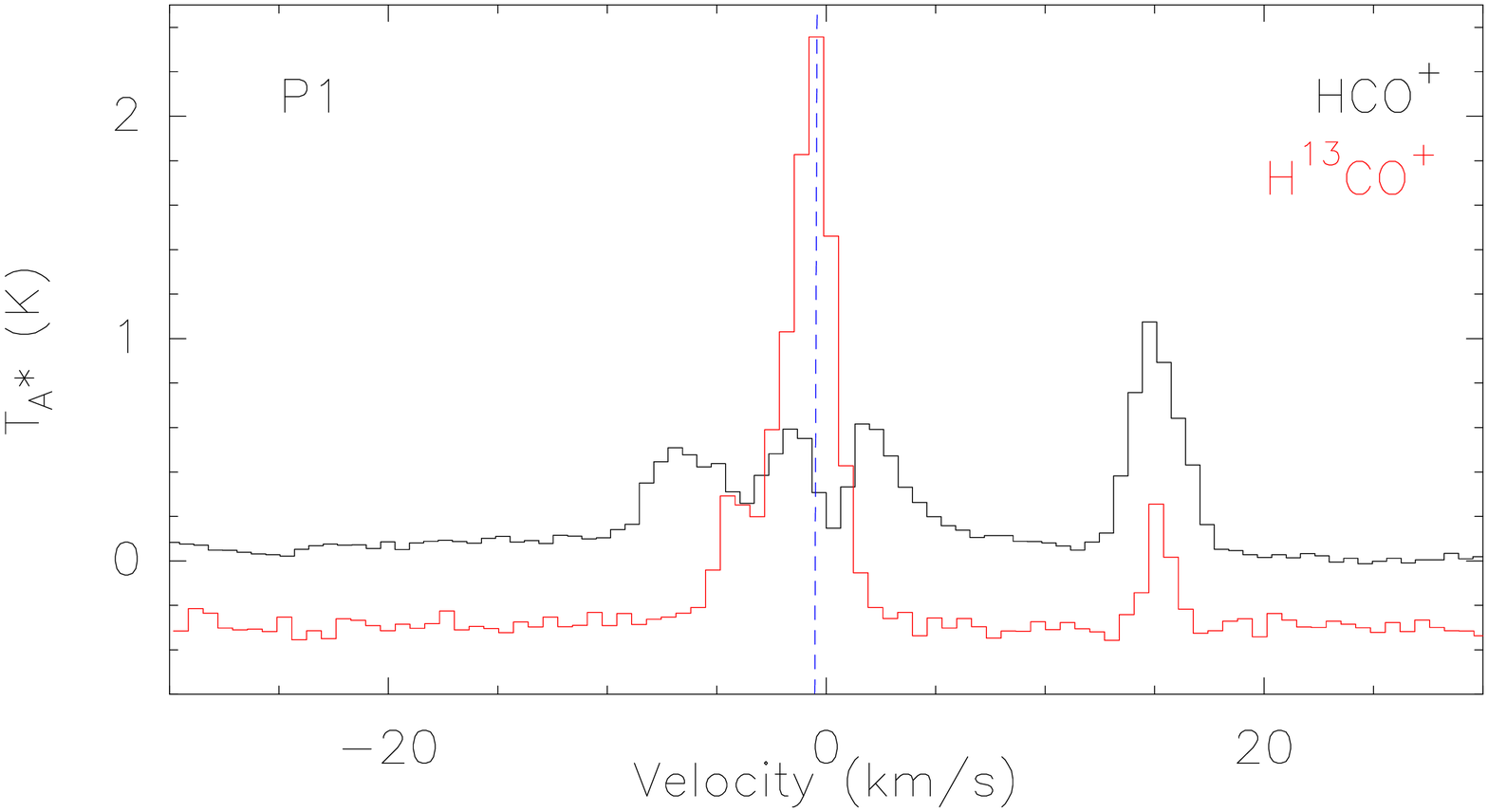} \quad \includegraphics[scale=0.16]{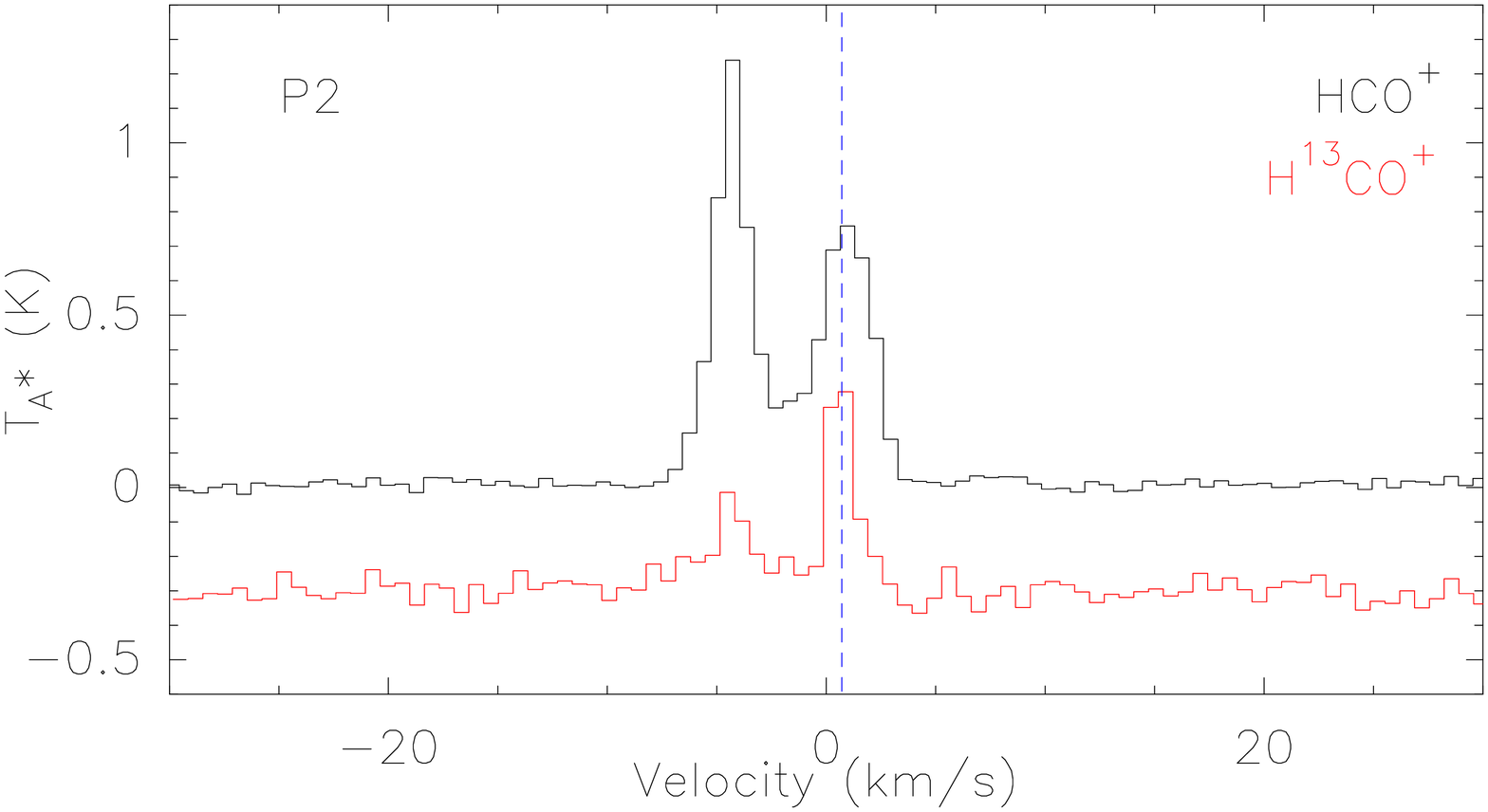} \quad \includegraphics[scale=0.16]{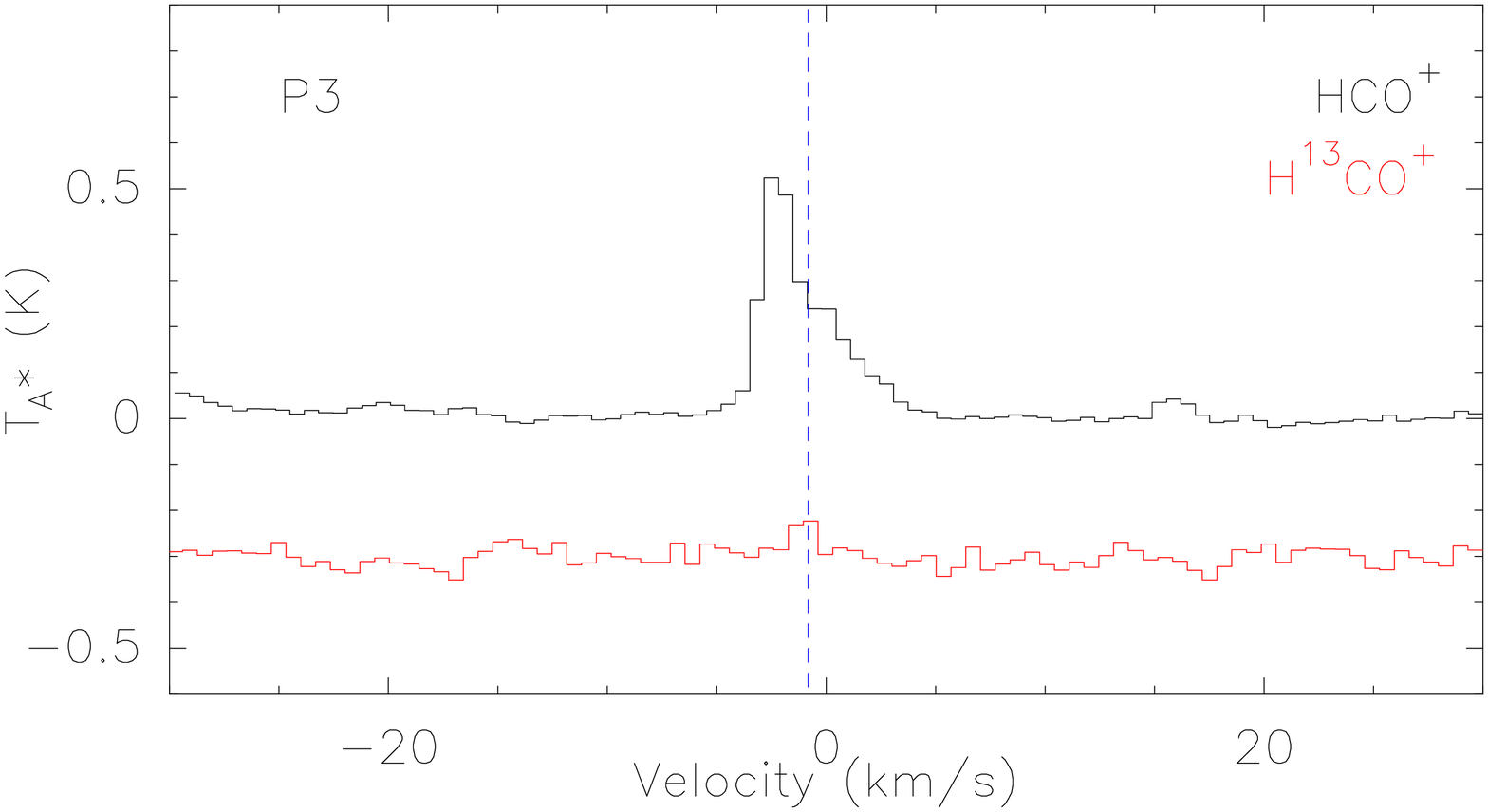} \quad \includegraphics[scale=0.16]{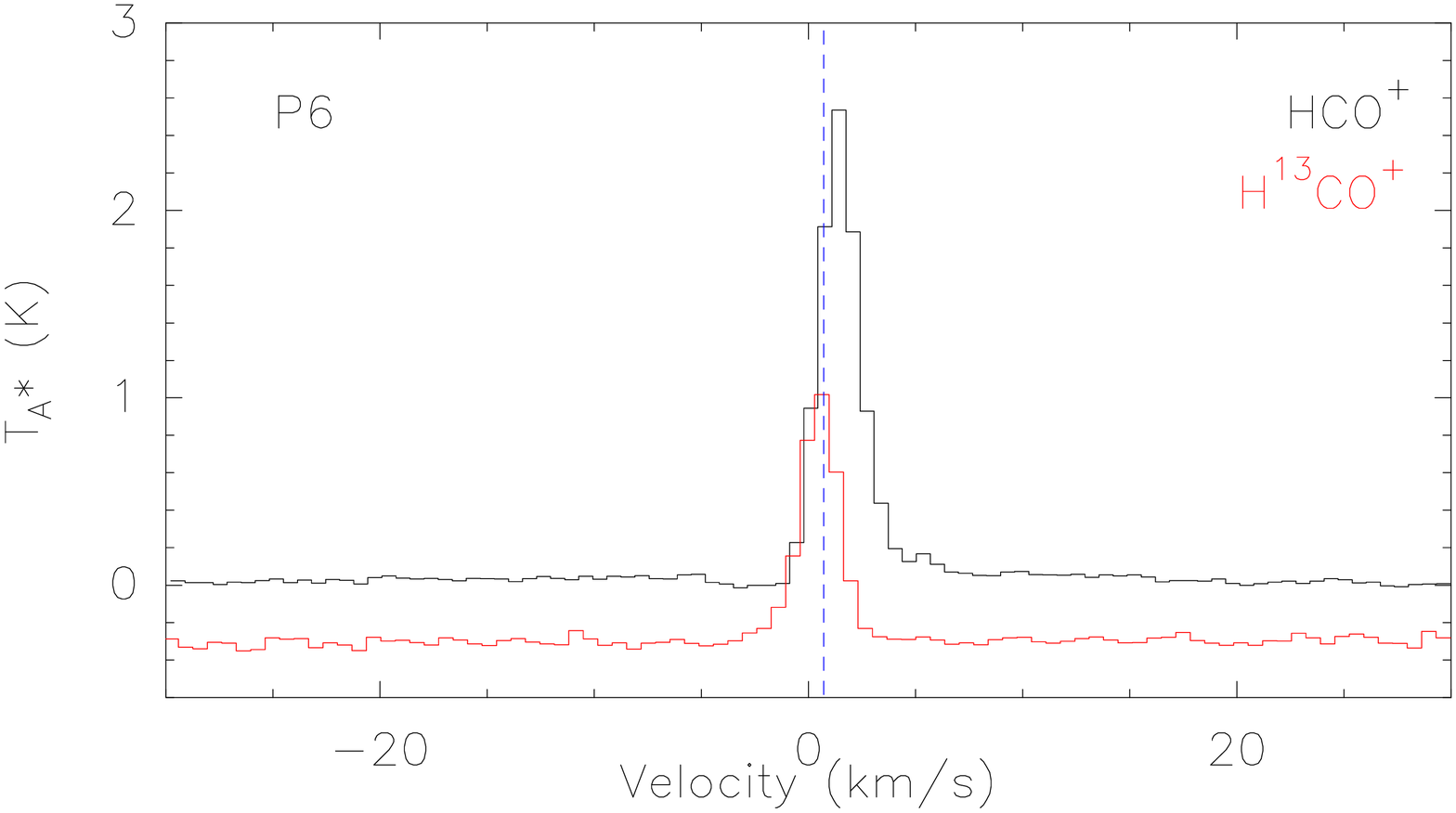} \quad \includegraphics[scale=0.16]{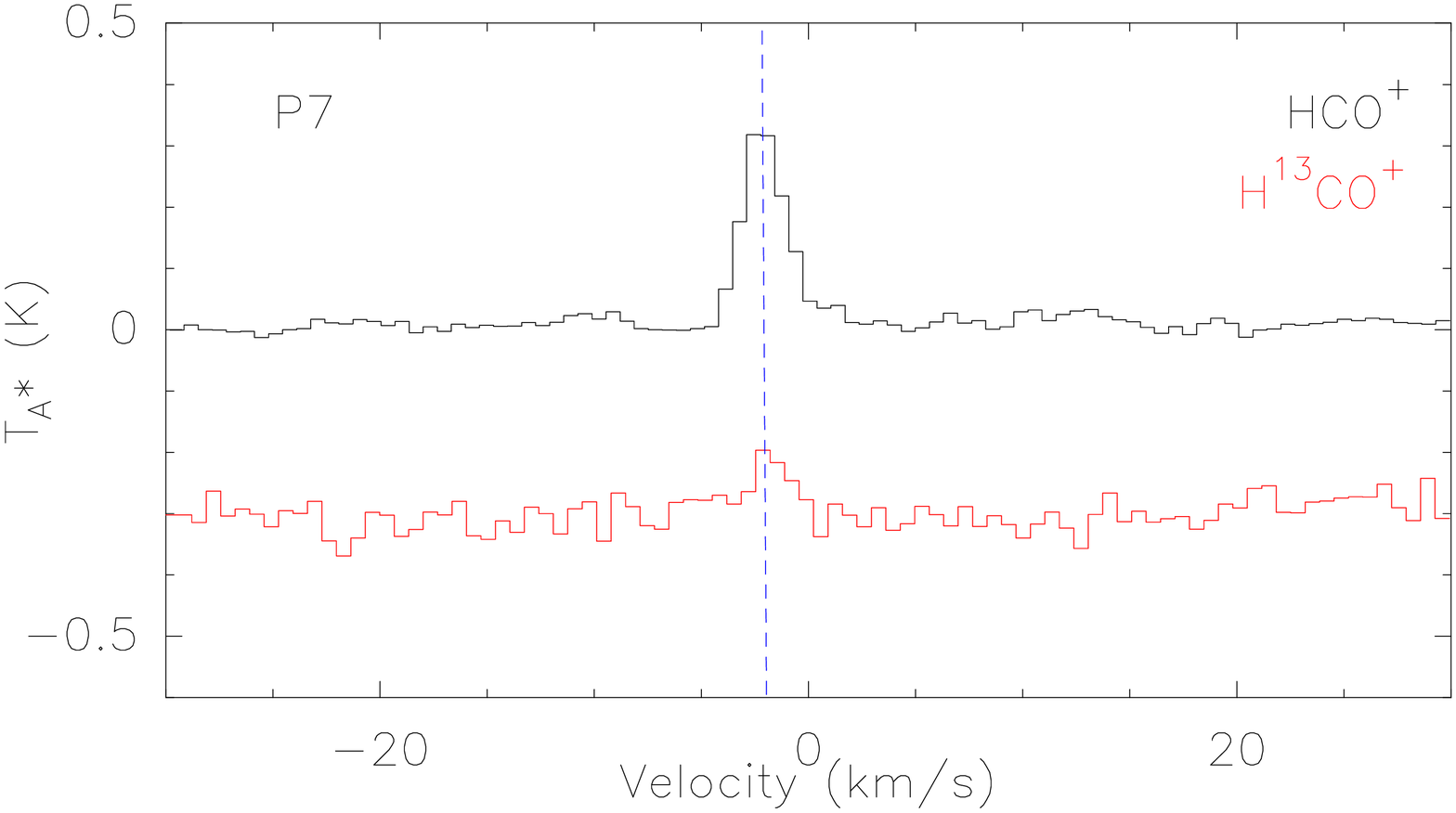} \quad \includegraphics[scale=0.16]{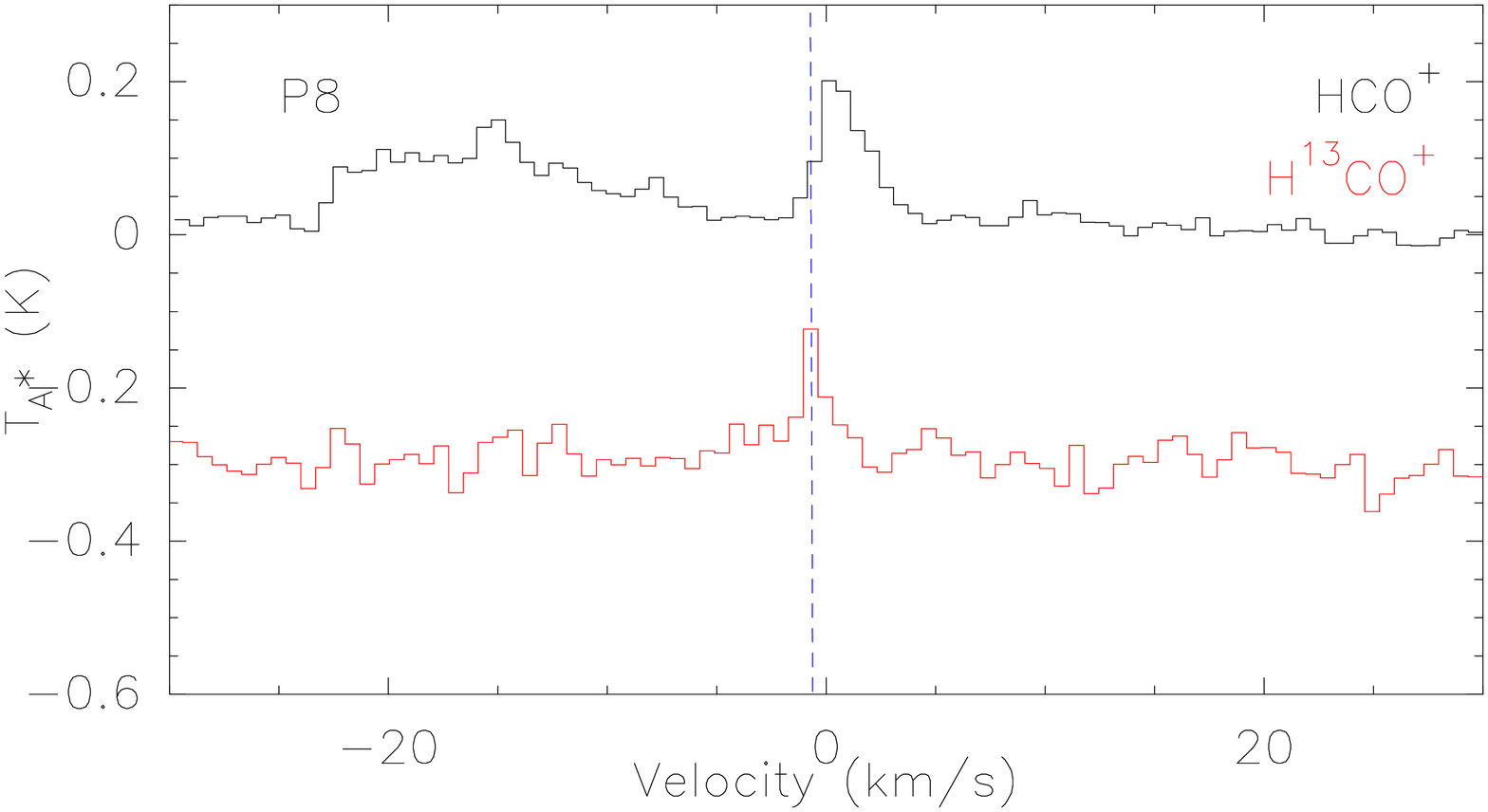} \quad \includegraphics[scale=0.16]{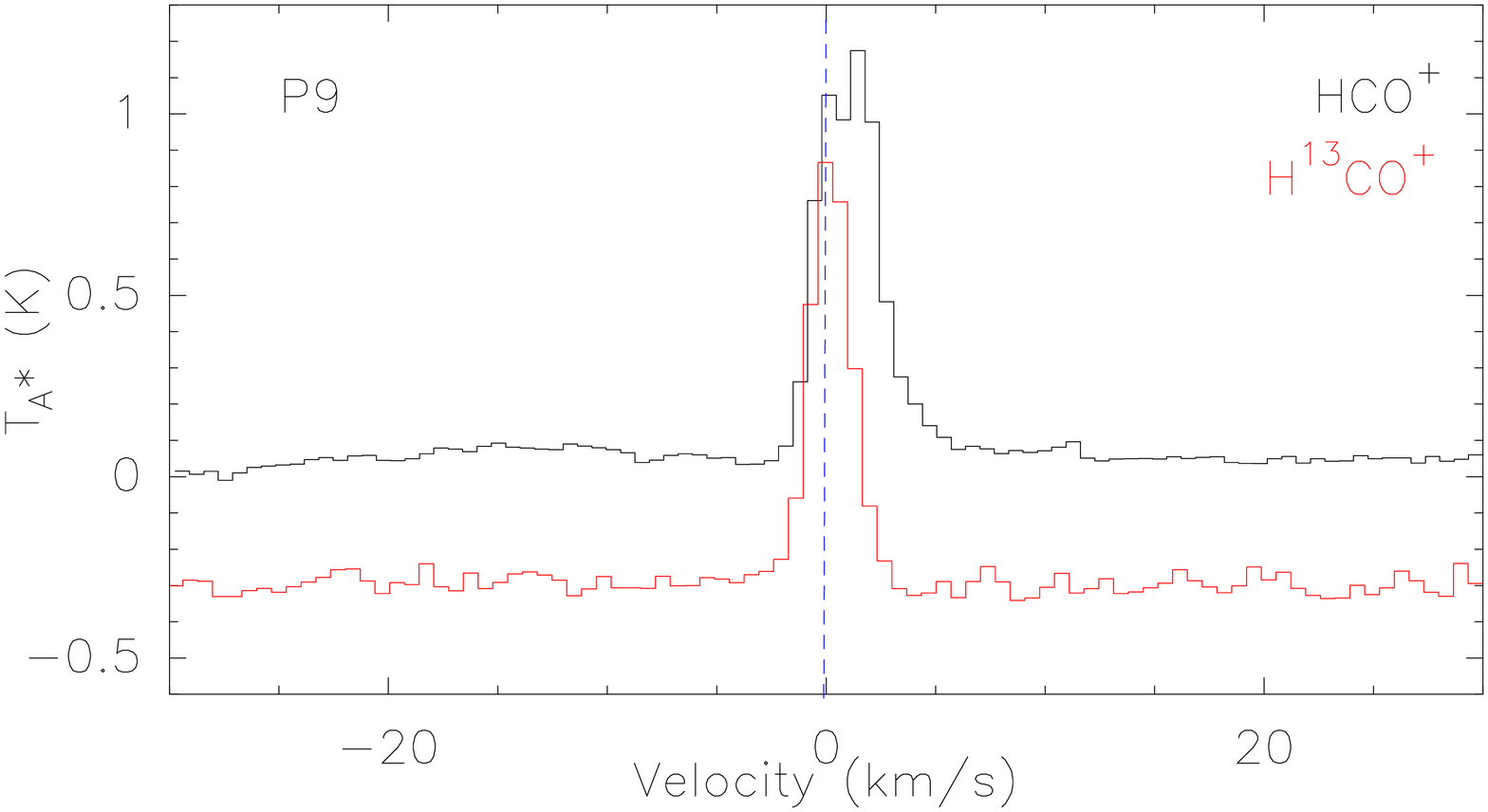} \quad \includegraphics[scale=0.16]{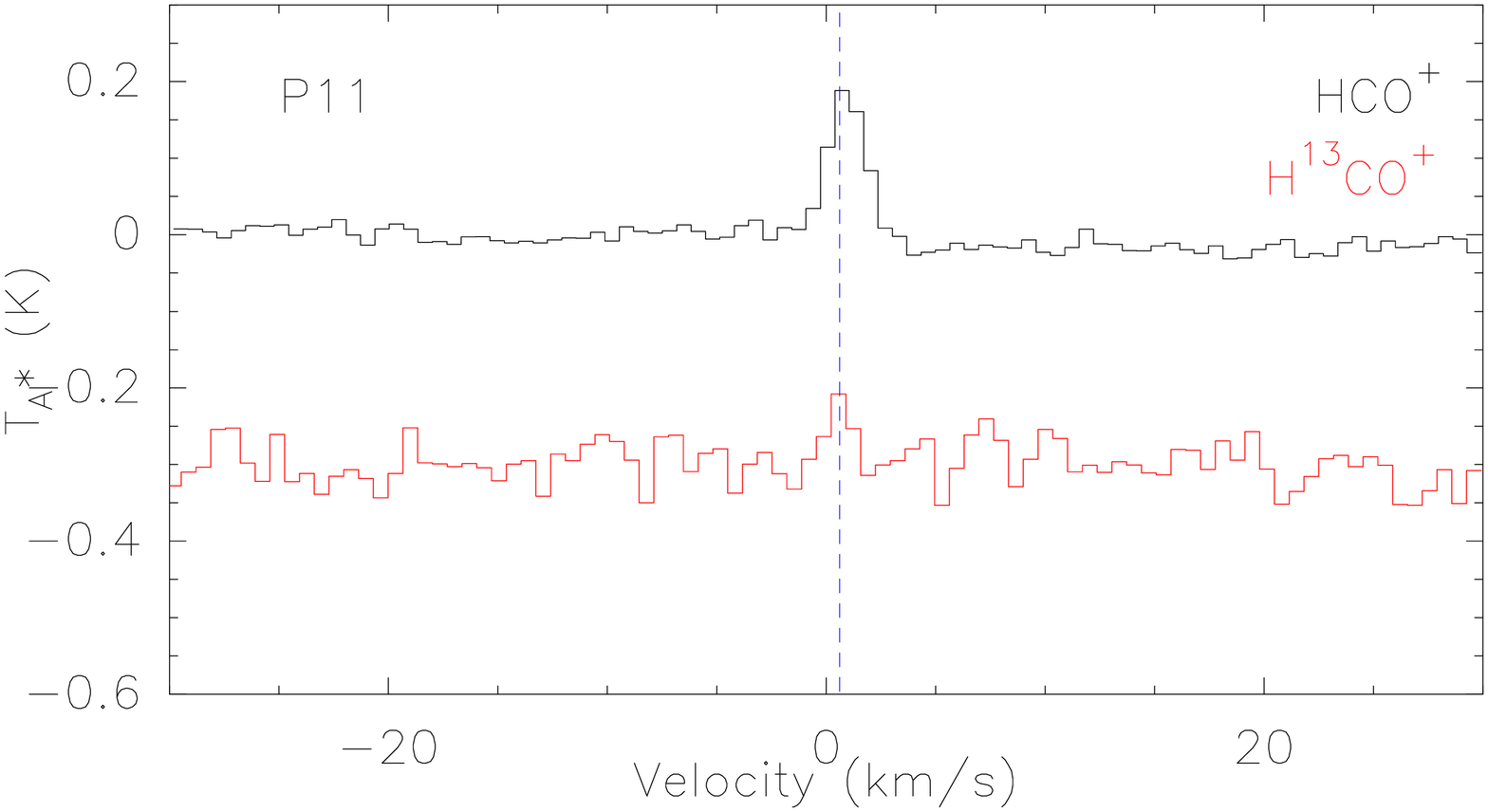}
\caption{Spectra of HCO$^+$ and H$^{13}$CO$^+$ $J=1-0$ emission toward 8 positions along the GCMF restricted to a velocity range of $-30$ to 30 km~s$^{-1}$ for a better visualisation of the emission features. $^{13}$C$^{18}$O spectra are scaled by a factor of 3. \textbf{Blue dashed line corresponds to peak of H$^{13}$CO$^+$ line.}}
\label{hcop}
\end{figure*}
\begin{figure*}[h!]
\centering
\hspace*{-0.6 cm}
\includegraphics[scale=0.16]{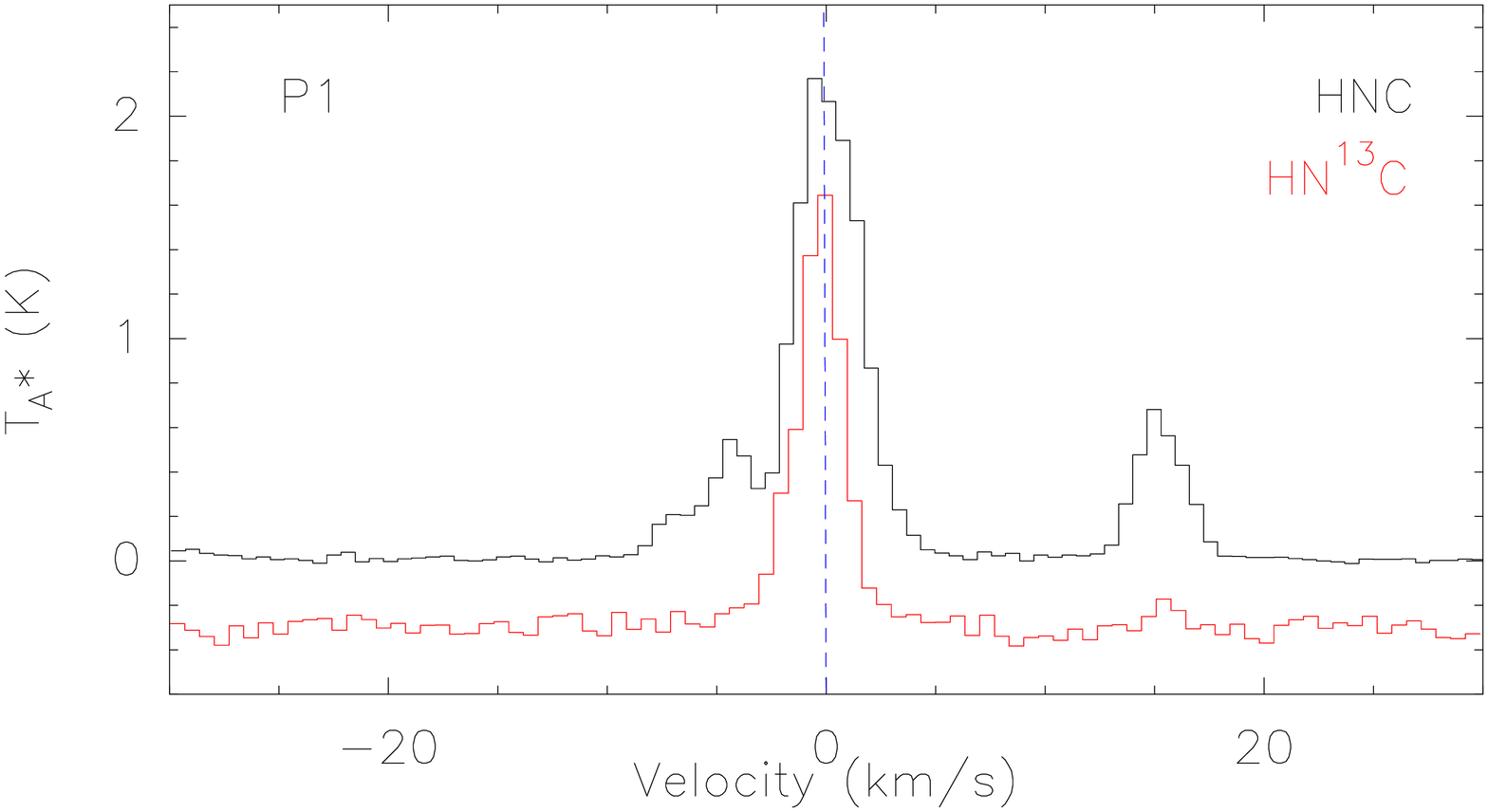}\quad \includegraphics[scale=0.16]{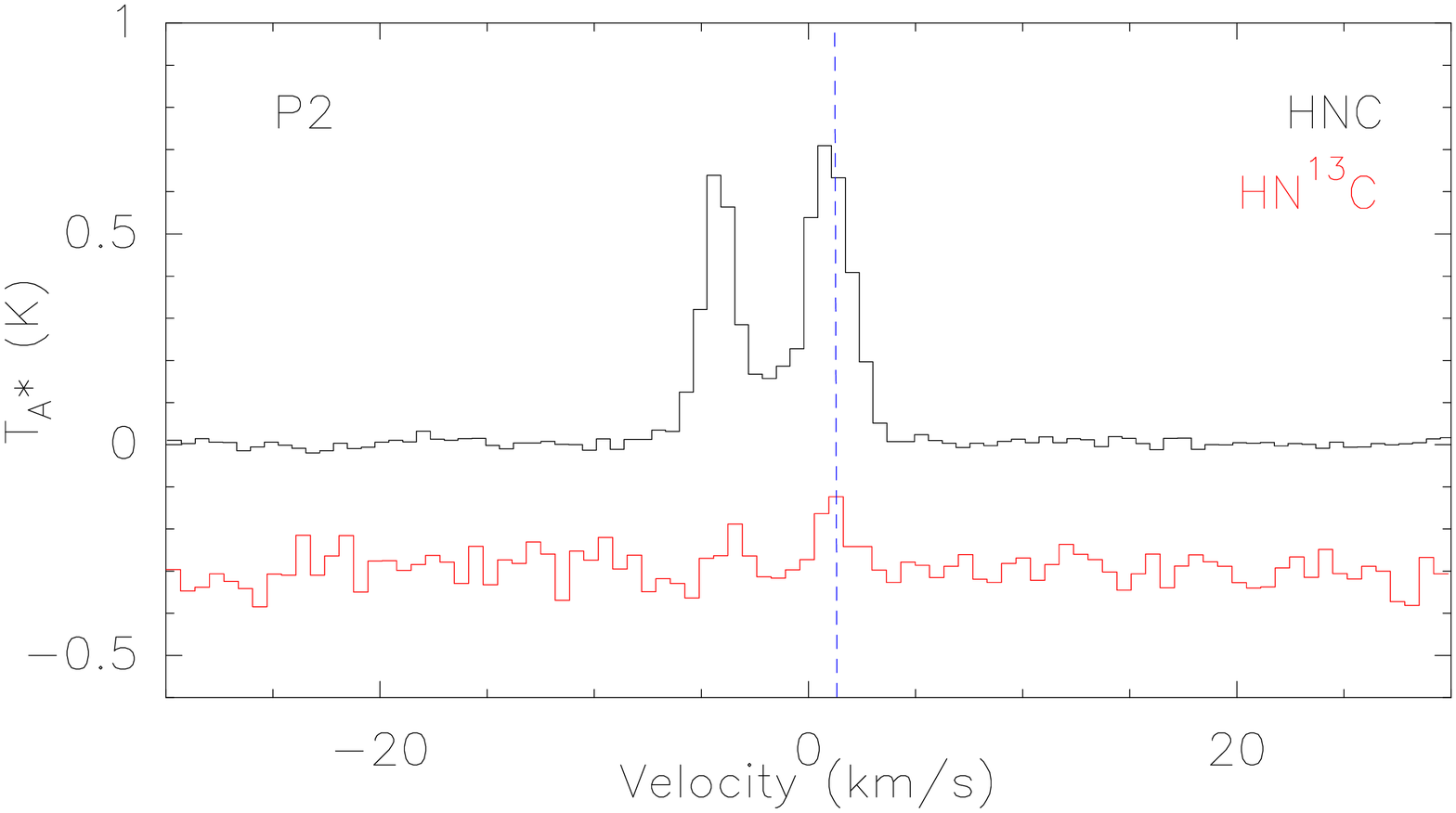} \quad \includegraphics[scale=0.16]{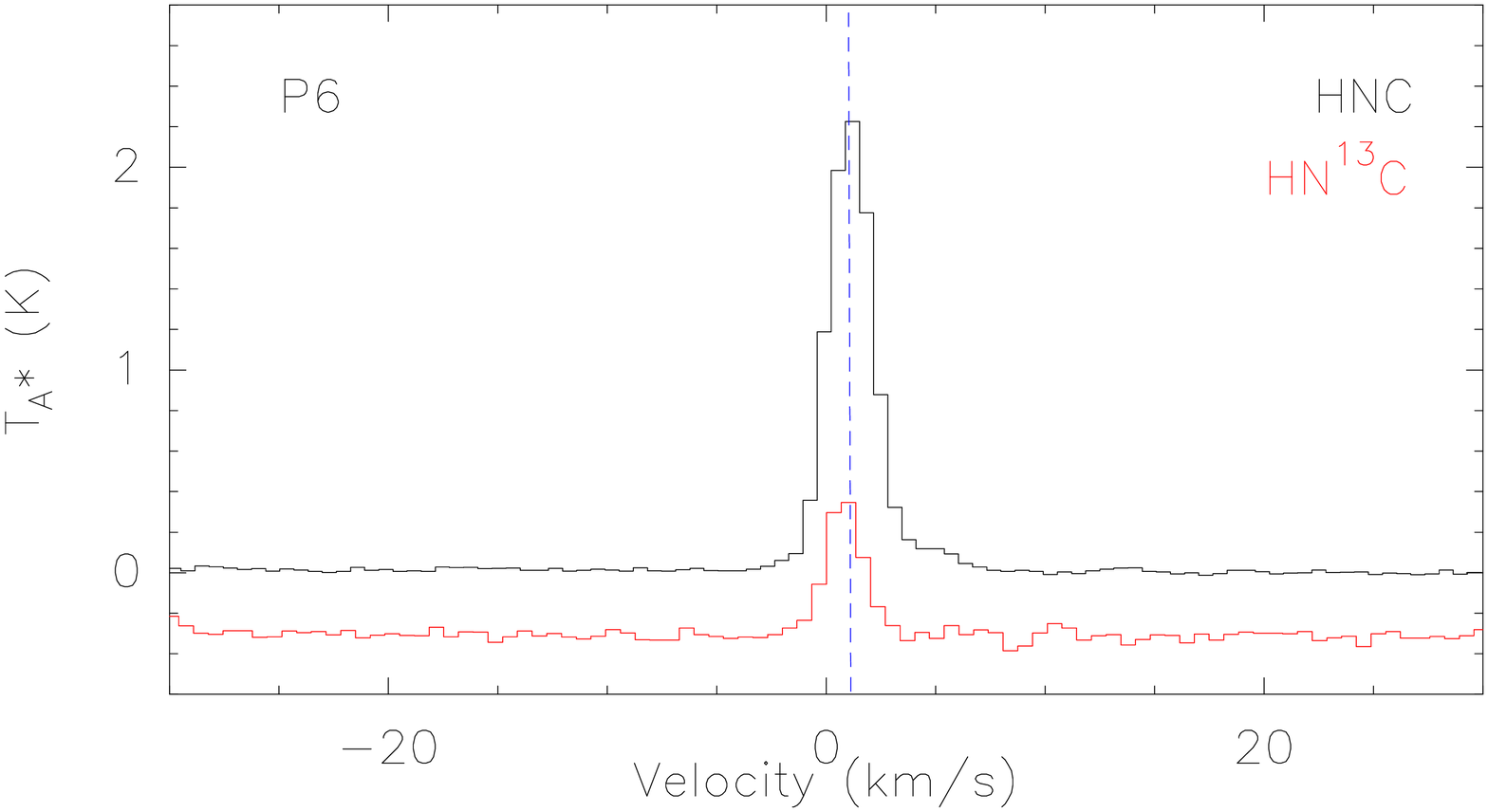} \quad \includegraphics[scale=0.16]{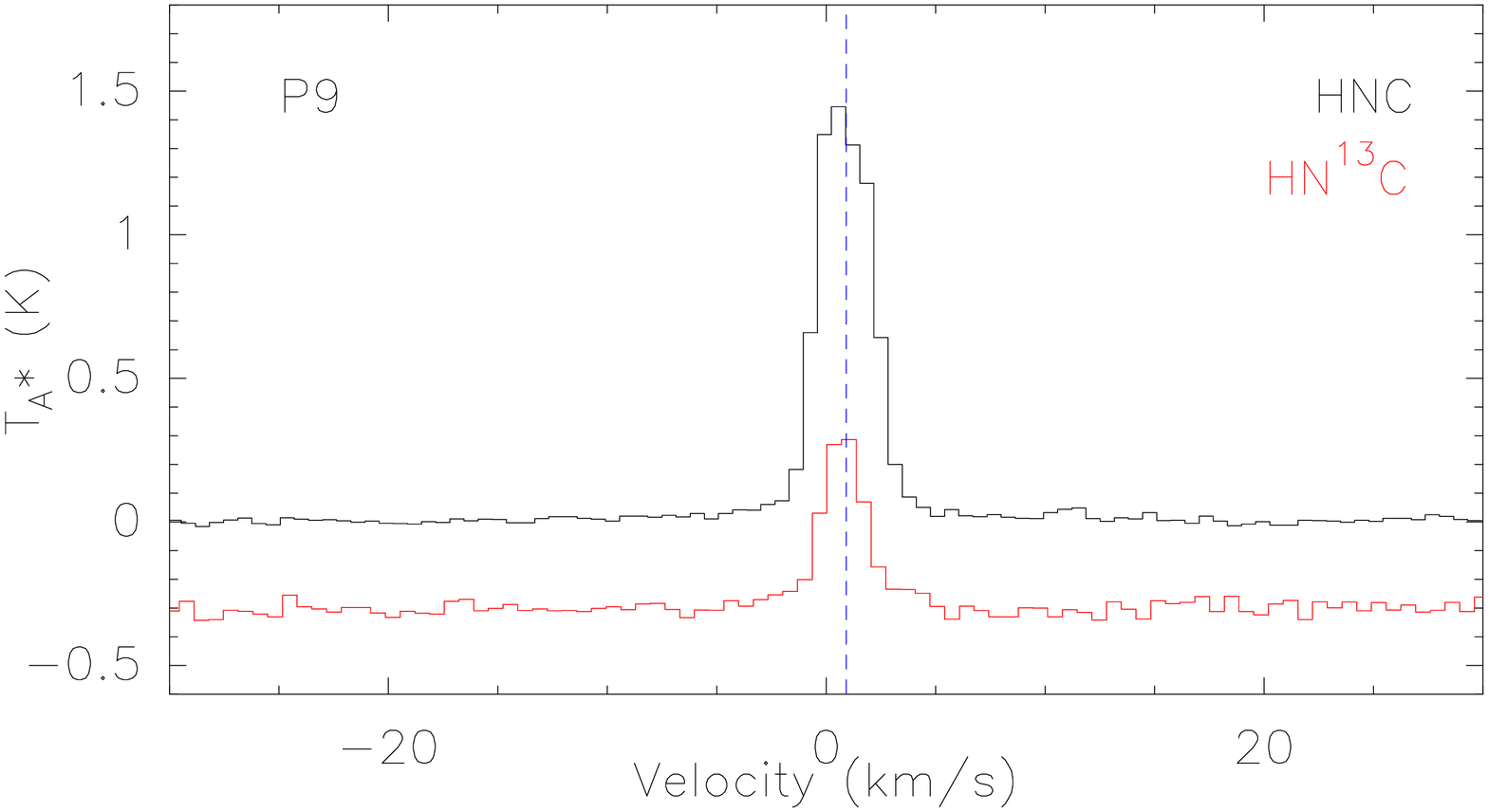}
\caption{Spectra of HNC and HN$^{13}$C $J=1-0$ emission toward 4 positions along the GCMF restricted to a velocity range of $-30$ to 30 km~s$^{-1}$ for a better visualisation of the emission features. $^{13}$CS spectra are scaled by a factor of 3. \textbf{Blue dashed line corresponds to peak of HN$^{13}$C line.}}
\label{hnc}
\end{figure*}
\begin{figure*}[h!]
\centering
\includegraphics[scale=0.5]{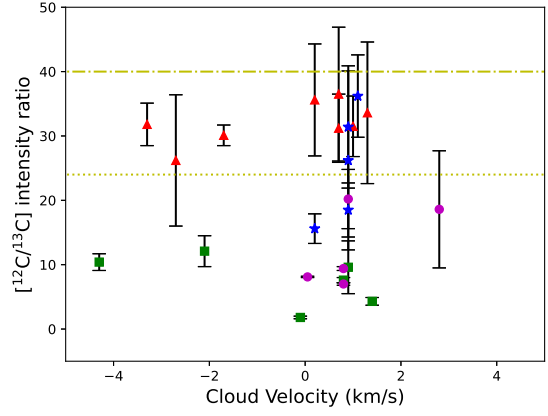} 
\caption{Zoomed-in view of Fig.~\ref{Cisograd}(Left) in the velocity range -5 to 5 km~s$^{-1}$ showing carbon isotope ratio of components belonging to GCMF.}
\label{Cisogradsub}
\end{figure*}

\begin{table}[!htbp]
\centering
\caption{Derived carbon isotope ratios towards GCMF based on C$^{18}$O and CS and their isotopologues.}
    \begin{tabular}{c c c}
 \hline
    \hline\\
Position&$\frac{\textrm I[\textrm C^{18}\textrm O]}{\textrm I[^{13}\textrm C^{18}\textrm O]}$&$\frac{\textrm{I[CS]}}{\textrm I[^{13}\textrm{CS}]}$\\
\hline
\hline \\
P1&$31.8\pm3.3$&$15.6\pm2.3$\\
P2&$31.5\pm4.7$&$31.4\pm9.5$\\
P3&$35.6\pm8.7$&-\\
P6&$36.5\pm10.4$&$36.2\pm6.4$\\
P7&$30.1\pm1.6$&-\\
P8&$33.6\pm11.0$&-\\
P9&$31.2\pm5.3$&$18.5\pm4.2$\\
P10&$26.2\pm10.2$&-\\
\hline\\
\end{tabular}
\label{isocsc18o}
\end{table}

\end{document}